\documentclass[fleqn,usenatbib]{mnras}

\usepackage{newtxtext,newtxmath}

\usepackage[T1]{fontenc}

\DeclareRobustCommand{\VAN}[3]{#2}
\let\VANthebibliography\thebibliography
\def\thebibliography{\DeclareRobustCommand{\VAN}[3]{##3}\VANthebibliography}

\usepackage{graphicx}	
\usepackage{amsmath}	
 
\usepackage{amssymb}	

\title[The cycle of metals in NGC\,1404]{The cycle of metals in the infalling elliptical galaxy NGC\,1404}

\author[F. Mernier et al.]{
F. Mernier,$^{1,2}$\thanks{E-mail: francois.mernier@esa.int}\thanks{\textit{ESA Research Fellow.}}
N. Werner,$^{3}$
Y. Su,$^{4}$
C. Pinto,$^{5}$
R. Grossov\'a,$^{3,6}$
A. Simionescu,$^{2,7,8}$
E. Iodice$^{9}$
\newauthor
M. Sarzi,$^{10}$ 
and A. G\"orgei$^{11}$
\\
$^{1}$ European Space Agency (ESA), European Space Research and Technology Centre (ESTEC), Keplerlaan 1, 2201 AZ Noordwijk, The Netherlands \\
$^{2}$SRON Netherlands Institute for Space Research, Niels Bohrweg 4, 2333 CA Leiden, The Netherlands\\
$^{3}$Department of Theoretical Physics and Astrophysics, Faculty of Science, Masaryk University, Kotl\'a\v{r}sk\'a 2, Brno, CZ-611 37, Czech Republic \\
$^{4}$Department of Physics and Astronomy, University of Kentucky, 505 Rose Street, Lexington, KY, 40506, USA \\
$^{5}$INAF - IASF Palermo, Via U. La Malfa 153, I-90146 Palermo, Italy \\
$^{6}$Dipartimento di Fisica, Universita degli Studi di Torino, via Pietro Giuria 1, I-10125 Torino, Italy \\
$^{7}$Leiden Observatory, Leiden University, PO Box 9513, NL-2300 RA Leiden, The Netherlands \\
$^{8}$Kavli Institute for the Physics and Mathematics of the Universe (WPI), University of Tokyo, Kashiwa 277-8583, Japan \\
$^{9}$INAF-Astronomical Observatory of Capodimonte, via Moiariello 16, I-80131, Napoli, Italy \\
$^{10}$Armagh Observatory and Planetarium, College Hill, Armagh BT61 DG, UK \\
$^{11}$Institute of Physics, E\"otv\"os University, P\'azm\'any P\'eter s\'et\'any 1/A, Budapest, 1117, Hungary
}

\date{Accepted 2022 January 24. Received 2022 January 14; in original form 2021 November 17}

\pubyear{2021}

\begin{document}
\label{firstpage}
\pagerange{\pageref{firstpage}--\pageref{lastpage}}
\maketitle

\begin{abstract}
Hot atmospheres pervading galaxy clusters, groups, and early-type galaxies are rich in metals, produced during epochs and diffused via processes that are still to be determined. While this enrichment has been routinely investigated in clusters, metals in lower mass systems are more challenging to probe with standard X-ray exposures and spectroscopy. In this paper, we focus on very deep \textit{XMM-Newton} ($\sim$350~ks) observations of NGC\,1404, a massive elliptical galaxy experiencing ram-pressure stripping of its hot atmosphere while infalling towards the centre of the Fornax cluster, with the aim to derive abundances through its hot gas extent. Importantly, we report the existence of a new fitting bias -- the ``double Fe bias'' -- leading to an underestimate of the Fe abundance when two thermal components cannot realistically model the complex temperature structure present in the outer atmosphere of the galaxy. Contrasting with the ``metal conundrum'' seen in clusters, the Fe and Mg masses of NGC\,1404 are measured 1--2 orders of magnitude \textit{below} what stars and supernovae could have reasonably produced and released. In addition, we note the remarkable Solar abundance ratios of the galaxy's halo, different from its stellar counterpart but similar to the chemical composition of the ICM of rich clusters. Completing the clusters regime, all these findings provide additional support towards a scenario of early enrichment, at play over two orders of magnitude in mass. A few peculiar and intriguing features, such as a possible double metal peak as well as an apparent ring of enhanced Si near the galaxy core, are also discussed.
\end{abstract}





\section{Introduction}
\label{sec:intro}


Over the two past decades, spatially resolved X-ray spectroscopy has considerably improved our knowledge of the growth of large-scale structures, such as galaxy clusters. The hot ($10^{7-8}$~K), X-ray emitting intracluster medium (ICM), falling into their deep gravitational well, constitutes in fact an unique way of probing the (thermo-) dynamics of most of clusters' baryons, and to understand their assembly over cosmic times \citep[for recent reviews, see e.g.][]{bohringer2010,simionescu2019b,werner2020}. 

Beyond long-standing questions over its thermal and assembly history, the ICM also has intriguing chemical properties. In fact, it is known for hosting a substantial fraction of metals, which must have been synthesized and released by supernovae (SNe), before enriching the entire cluster volume \citep[for recent reviews, see e.g.][]{biffi2018b,mernier2018c}. As clusters can retain all baryons permanently, the chemical elements they accumulate constitute a remarkable fossil record of the bulk enrichment of our Universe. More specifically: $\alpha$-elements (e.g. O, Ne, Mg, Si, S) are mainly produced by core-collapse supernovae (SNcc), Fe-peak elements (e.g. Ca, Cr, Mn, Fe, Ni) are mainly produced by Type Ia supernovae (SNIa), while asymptotic giant branch (AGB) stars produce and release most of C and N elements \citep[for a review, see e.g.][]{nomoto2013}. As the aforementioned elements are all accessible via their K-shell (sometimes L-shell) emission lines in the energy band of the currently flying X-ray observatories, one can study the spatial distribution, relative ratios, and redshift-evolution of their abundances in order to better understand (i) the cosmic history of metals, (ii) their share and budget between cluster components, and (iii) their transport and diffusion mechanisms from sub-pc (SNe) to Mpc scales (clusters and groups).

Since the beginning of the \textit{XMM-Newton} and \textit{Chandra} era, many essential discoveries were done in this respect. Among them, one can certainly cite:

\begin{itemize}

\item The presence of a central Fe peak in the most relaxed systems \citep[e.g.][]{degrandi2001}, with a very similar distribution for the other elements \citep[e.g.][]{deplaa2006,simionescu2009,million2011,mernier2017};

\item A remarkably uniform metallicity distribution in cluster outskirts, sometimes out to their virial radius \citep[][]{werner2013,urban2017,ghizzardi2021};

\item A chemical composition very close to that of our own Solar System \citep{deplaa2007,degrandi2009,mernier2016a,mernier2016b,hitomi2017,mernier2018b,simionescu2019a};

\item Redshift studies consistent with no evolutionary trend of abundances out to at least $z \sim 1.5$ \citep[though with non-negligible uncertainties;][]{ettori2015,mcdonald2016,mantz2017,liu2018,flores2021};

\item The ICM being too metal-rich compared to what could be reasonably produced by the stellar population in cluster galaxies, leading to a potentially serious conundrum in the metal budget at large scales \citep{arnaud1992,bregman2010,loewenstein2013,renzini2014}.

\end{itemize}

Together with cosmological simulations \citep[e.g.][]{biffi2017}, these observations progressively gathered key evidence towards a scenario in which the ICM completed its enrichment at or beyond $z \sim 2$--3  -- coinciding with the main epoch of peak cosmic star formation history and maximal supermassive black hole activity \citep{madau2014,hickox2018} -- with metals being thoroughly ejected and mixed at Mpc scales via feedback from active galactic nuclei (AGNs).

The case of lower mass systems, i.e. galaxy groups and early-type galaxies (ETGs), has been less studied and, consequently, less understood \citep[for a recent review, see][]{gastaldello2021}. The intragroup medium (IGrM) and the hot, X-ray emitting atmosphere pervading elliptical galaxies are also rich in metals, detected essentially via the Fe-L complex; however the latter remains essentially unresolved at current CCD spectral resolution, making the Fe abundance more sensitive to systematics. In addition, these systems are typically 1--2 order(s) of magnitude less bright than their cluster counterparts, hence requiring deeper observations. Interestingly, metal peaks are also observed in groups \citep[e.g.][]{sun2012,mernier2017,lovisari2019} and, though less investigated than clusters, the metal mass in the IGrM tends to better agree with the available production from stellar sources \citep[e.g.][]{renzini2014,sasaki2014}. The case of these lower mass systems, however, is more challenging to interpret, as their shallower gravitational well (resulting in relatively stronger AGN feedback, competing with external gas accretion) prevents them from being considered as closed-box systems. Also, despite growing indications towards a chemical composition not so different between the ICM and the IGrM \citep{kim2004,loewenstein2010,loewenstein2012,mernier2018b}, the more extreme case of ETGs needs to be more thoroughly addressed. Ultimately, linking the chemical budget and history of ETGs (and groups) to that of clusters is absolutely essential to unify the enrichment picture at \textit{all} extragalactic scales.

An excellent target to tackle these issues is NGC\,1404 (FCC\,219). This massive elliptical galaxy \citep[$12.7 \times 10^{10}$~$M_\odot$;][]{iodice2019} is falling towards the centre of the Fornax cluster (particularly its central dominant galaxy NGC\,1399, a.k.a. FCC\,213), and its relative proximity (less than 20 Mpc) allows its gas properties to be studied in great detail. Though likely at its second or third passage already \citep{sheardown2018}, its infall makes it a textbook example of ram-pressure stripping at play between a galaxy's hot atmosphere and its surrounding ICM \citep{machacek2005,su2017a,su2017b,su2017d}. The particular interaction of this ETG with the Fornax cluster makes its dynamics essentially restricted to (i) gas loss (via ram-pressure stripping) and (ii) a discontinuous interface (cold front) between the two media, in which Kelvin-Helmholtz instabilities may develop \citep[depending on the gas viscosity and magnetic draping;][]{su2017b}; however, it also prevents its hot atmosphere from experiencing substantial external accretion. Last but certainly not least, this galaxy does \textit{not} have any known AGN activity (i.e. neither strong nuclear nor diffuse radio emission, nor X-ray cavities have been detected so far), and there are interesting indications of its atmosphere being rather turbulent compared with that of other ETGs \citep{pinto2015,ogorzalek2017}.

\begin{table*}
\begin{centering}
\caption{List of \textit{XMM-Newton} and \textit{Chandra} observations used in this work. A fifth \textit{XMM-Netwon} pointing (ObsID:0055140101) is not used here, as the position of NGC\,1404 close to the detector edge (hence strongly impacting the PSF of the source) would have considerably complicated our analysis. $^{\mathrm{a}}$ RGS data presented in this work are analysed from this pointing only. $^{\mathrm{b}}$ These pointings do not cover the centre of NGC\,1404 and, therefore, are used for imaging purposes only (Fig.~\ref{fig:NGC1404}). $^{\mathrm{c}}$ Includes only observations covering NGC\,1404 (hence suitable for spectroscopy analysis; see the text).}             
\label{tab:observations}
\begin{tabular}{l l c c c}        
\hline \hline                
Observatory & ObsID & Observation date & Raw exposure & Net exposure \\
 &  & (yyyy-mm--dd) & (ks) & (ks) \\    
\hline
\textit{XMM-Newton}	& 0012830101 & 2001-06-27 & 29.2 & 6.3 \\
				& 0304940101 & 2005-07-30 & 55.0 & 28.4 \\
				& 0400620101 & 2006-08-23 & 130.0 & 114.1 \\
				& 0781350101$^{\mathrm{a}}$ & 2016-12-29 & 134.3 & 126.7 \\
				\hline
				& Total & 		-		        & 348.5 & 275.5 \\
				\hline
\textit{Chandra}	& 240$^{\mathrm{b}}$ & 2000-06-16 & 43.5 & 44.1 \\
(ACIS-S)			& 319 & 2000-01-18 & 60.3 & 56.8 \\
				& 2389$^{\mathrm{b}}$ & 2001-05-08 & 14.7 & 14.7 \\
				& 2942 & 2003-02-13 & 30.0 & 29.6 \\
				& 9530$^{\mathrm{b}}$ & 2008-06-08 & 65.0 & 60.1 \\
				& 9798 & 2007-12-24 & 18.3 & 18.3 \\
				& 9799 & 2007-12-27 & 21.3 & 21.3 \\
				& 14527 & 2013-07-01 & 27.8 & 27.8 \\
				& 14529 & 2015-11-06 & 31.6 & 31.6 \\
				& 16231 & 2014-10-20 & 60.5 & 60.5 \\
				& 16232 & 2014-11-12 & 69.1 & 69.1 \\
				& 16233 & 2014-11-09 & 98.8 & 98.8 \\
				& 16639 & 2014-10-12 & 29.7 & 29.7 \\
				& 17540 & 2016-04-02 & 28.5 & 28.5 \\
				& 17541 & 2014-10-23 & 24.7 & 25.1 \\
				& 17548 & 2014-11-11 & 48.2 & 48.2 \\
				& 17549 & 2015-03-28 & 61.6 & 61.6 \\
				\hline
\textit{Chandra}	& 239 & 2000-01-19 & 3.6 & 3.6 \\
(ACIS-I)			& 4172$^{\mathrm{b}}$ & 2003-05-26 & 44.5 & 44.5 \\
				& 4174 & 2003-05-28 & 50.0 & 46.3 \\
				\hline			
				& Total$^{\mathrm{c}}$ & 		-		        & 664.0 & 656.8 \\
\hline                                   
\end{tabular}
\par\end{centering}
\end{table*}

Thanks to its peculiar dynamics, NGC\,1404 has benefited from deep \textit{Chandra} exposures, so far mostly used for imaging purposes \citep[][]{su2017a,su2017b}. Recent deep \textit{XMM-Newton} observations were also conducted to derive high upper limits on its turbulence \citep{bambic2018}. In both cases, however, the chemical enrichment aspect of its X-ray halo has not been addressed in detail. Accordingly, in this paper, we take advantage of these observations to probe the metal distribution of NGC\,1404 as well as its chemical composition with unprecedented accuracy. The emphasis is put on \textit{XMM-Newton} for its spectroscopic capabilities, via its three European Photon Imaging Cameras (EPIC; allowing spatial spectroscopy at moderate energy resolution) and its two Reflection Grating Spectrometers (RGS; allowing high resolution spectroscopy across a central detector strip). Nevertheless, we also show how data from the Advanced CCD Imaging Spectrometer (ACIS) onboard \textit{Chandra} are nicely complementary in specific cases. 

Throughout this paper, we assume a $\Lambda$CDM cosmology with $H_0 = 70$~km~s$^{-1}$~Mpc$^{-1}$, $\Omega_m = 0.3$, and $\Omega_\Lambda = 0.7$. We assume NGC\,1404 to be at a distance of 18.79~Mpc \citep{tully2016}, for which 1 arcmin corresponds to 5.47 kpc. Its optical radii $R_e$ and $R_{25}$ \citep[respectively 2.3~kpc and 11.5~kpc;][]{sarzi2018} are fully covered by our spatially resolved analysis. Following \citet{pinto2015}, we adopt $r_{500} = 0.61$~Mpc. All the abundances are referred with respect to the proto-solar values of \citet[][referred to as "Solar" for simplicity]{lodders2009}. Unless stated otherwise, the error bars correspond to the 68\% confidence level.


\section{Data reduction and analysis}\label{sec:data_reduction}


\begin{figure*}
        \centering
                \includegraphics[width=\textwidth, trim={0cm 0cm 0cm 1.8cm},clip]{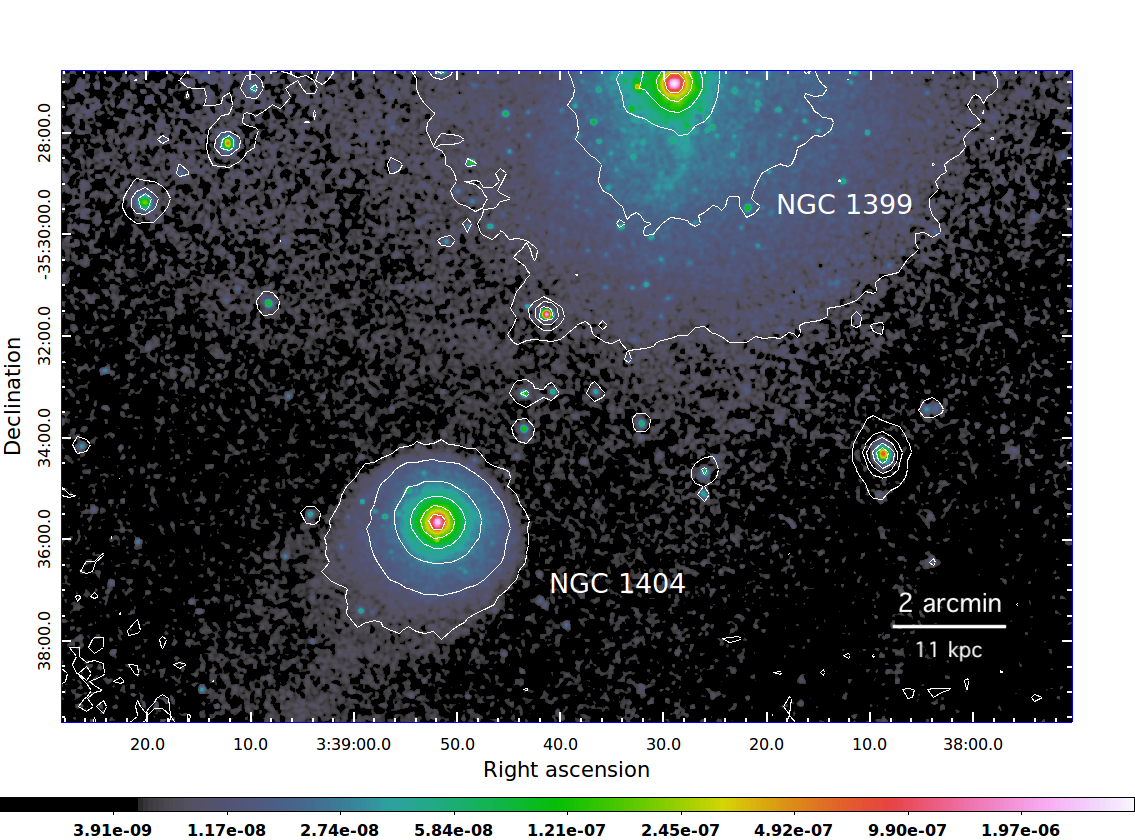}
        \caption{Adaptively smoothed, exposure-corrected flux map of NGC\,1404 (and NGC\,1399) observed from combined \textit{Chandra}/ACIS observations in the 0.3--2 keV band. This mosaic includes ACIS observations that do not necessarily cover the immediate surroundings of NGC\,1404 (see also Table~\ref{tab:observations}). The white contours correspond to the background-, exposure-corrected \textit{XMM-Newton}/EPIC counts (MOS\,1 + MOS\,2 + pn) in the same energy band.}
\label{fig:NGC1404}
\end{figure*}

\begin{figure}
        \centering
                \includegraphics[width=0.49\textwidth, trim={0cm 0.1cm 0cm 0cm},clip]{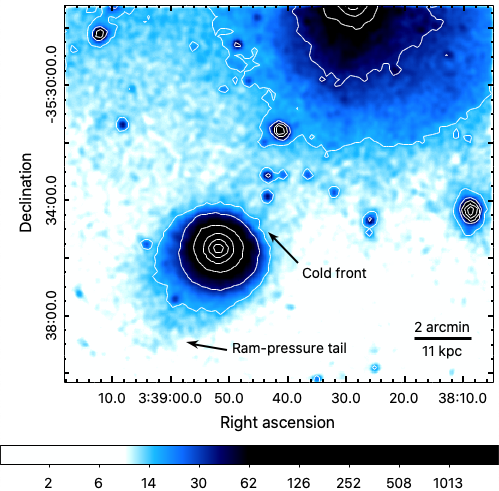}
        \caption{Exposure-corrected \textit{XMM-Newton}/EPIC counts image of NGC\,1404 in the 0.3--2~keV band, with its corresponding SB contours. The ram-pressure tail in the SE direction, as well as the NW merging cold front, are indicated.}
\label{fig:NGC1404_xmm}
\end{figure}

\subsection{\textit{Chandra} observations}\label{sec:data_chandra}

Table~\ref{tab:observations} lists the 20 archival \textit{Chandra} pointings with the ACIS-S or ACIS-I chips covering NGC\,1404 and/or its immediate surroundings. We reduce all these data using the \texttt{CIAO} (v4.11) analysis package as well as the calibration files of 2019 September (CALDB v4.8.4.1). Following the standard procedure, we reprocess the data reduction using the \texttt{CIAO} task \texttt{chandra\_repro} before filtering them from possible soft proton flares in the 0.5--7 keV band using the 2$\sigma$ clipping setup of the task \texttt{deflare}. The images are then co-added and exposure-corrected in the 0.3--2 keV band (via the routine \texttt{merge\_obs}), before we adaptively smooth the combined result (via the algorithm \texttt{csmooth}). The resulting flux map, shown in Fig.~\ref{fig:NGC1404}, reveals some exquisite features of the hot atmosphere of NGC\,1404 during its interaction with the ambient ICM surrounding NGC\,1399 \citep[e.g. the well known cold front and ram-pressure tail;][]{machacek2005,su2017a,su2017b,su2017d}. We also carefully identify point-sources irrelevant to our analysis (either behind or belonging to NGC\,1404), by using the task \texttt{wavdetect}. Finally, the raw spectra and their associated the response matrix files (RMF) and ancillary response files (ARF) are generated via the task \texttt{specextract}.

\subsection{\textit{XMM-Newton} observations}\label{sec:data_xmm}

The bulk of this work relies on \textit{XMM-Newton} observations of NGC\,1404, as we aim to take advantage of the excellent spectral capabilities and effective area of its EPIC and RGS instruments. We use the four observations that are publicly available in the archive, as listed in Table~\ref{tab:observations}, and reduce them using the \textit{XMM-Newton} Science Analysis System (\texttt{SAS} v17.0.0). The calibration files of 2019 January are used.

\subsubsection{EPIC}

Following the general prescription of \citet{mernier2015}, the MOS (i.e. MOS\,1 and MOS\,2) and pn data are pre-processed using the task \texttt{emproc} and \texttt{epproc}, respectively. We keep the single, double, triple, and quadruple events in the MOS data (\texttt{pattern$\le$12}) and restrict the pn data to single events only (\texttt{pattern=0}). To minimise the detector contamination by soft protons emitted by solar flares, we use the \texttt{SAS} task \texttt{espfilt} to define good time intervals (GTIs) that are then used to filter our data. Specifically, light curves in the 10--12 keV band are used to derive 100\,s count rate histograms. Count rates exceeding 2$\sigma$ from the mean of the best-fitting (Gaussian) distribution are then excluded from our GTIs. More accurate than selecting count rate thresholds arbitrarily, this 2$\sigma$ clipping method is also repeated in the 0.3--2 keV band, as flares were reported at soft energies as well \citep{deluca2004}. We also use the task \texttt{edetect\_chain} to detect and further exclude point sources. Images and raw spectra are then extracted using the task \texttt{evselect}. A combined, exposure-corrected image (co-added with the task \texttt{emosaic}) focusing on the ram-pressure tail is shown in Fig.~\ref{fig:NGC1404_xmm}. The RMF and ARF are obtained using the tasks \texttt{rmfgen} and \texttt{arfgen}, respectively.

\subsubsection{RGS}\label{sec:data_rgs}

Among the four available \textit{XMM-Newton} pointings, only two (of 134~ks and 55~ks) are centred on NGC\,1404 and are thus suitable for RGS analysis. Among these two, we choose to rely only on the RGS data available taken with the deepest \textit{XMM-Newton} observation (ObsID:0781350101 -- see Table~\ref{tab:observations}). This choice is made for self-consistency and to avoid unnecessary difficulties in the analysis, since the two observations are made at different roll-angles and their central coordinates slightly differ as well. Such risk for extra systematic effects would be quite important in comparison to the low ($\lesssim$20\%) statistics improvement expected from co-adding that second pointing.

The data are extracted and obtained following the same procedure as in \citet{pinto2015} and \citet{mernier2015}, i.e. based on the \texttt{SAS} task \texttt{rgsproc}. During the process, we also filter RGS\,1 and RGS\,2 data from flaring events using the same GTIs as for MOS\,1 and MOS\,2, respectively. We choose to include 90\% of the point spread function (PSF) along the cross-dispersion direction (\texttt{xpsfincl=90}), which represents a spatial width of about 0.8 arcmin. This choice allows us to minimise the instrumental line broadening (due to the slit-less feature of the gratings -- see Sect. \ref{sec:analysis}) while keeping a large number of counts (thanks to the high central flux of the source combined with the deep exposure of our selected observation). We then combine the RGS\,1 and RGS\,2 spectral data and responses for each order separately using the \texttt{SPEX} task \texttt{rgscombine}. These combined spectra are fitted simultaneously for order 1 and 2.

\begin{figure}
        \centering
                \includegraphics[width=0.455\textwidth, trim={0cm 0cm 0cm 0.1cm},clip]{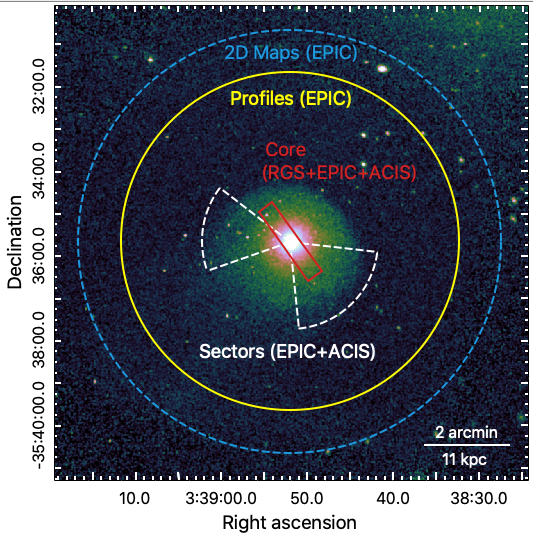}
        \caption{Zoomed-in \textit{Chandra}/ACIS image of NGC\,1404 (Fig.~\ref{fig:NGC1404}). The spectral extraction regions used in this work are indicated and their choice are further motivated in the text (Sect.~\ref{sec:data_regions}). The central rectangular region (red) coincides with the RGS coverage of the deepest \textit{XMM-Newton} observation, selected with $\sim$0.8 arcmin of cross-dispersion width.}
\label{fig:NGC1404_zoomed}
\end{figure}

\subsection{Selected regions}\label{sec:data_regions}

As we discuss further in this paper, NGC\,1404 and its chemical enrichment are interesting on many aspects. Figure~\ref{fig:NGC1404_zoomed} summarizes the spatial locations that are considered in this analysis. These locations include (i) a box-shaped region covering the galaxy core, (ii) azimuthally averaged radial profiles, and (iii) 2D maps of the galaxy and its vicinity.

The ``core'' (i.e. central box) region is based on the extraction region of the deepest RGS pointing (Sect.~\ref{sec:data_rgs}). The box is defined such that its length direction is aligned with the RGS dispersion direction of the latter. Although this work relies essentially on the spectroscopic capabilities of the EPIC and RGS instruments onboard \textit{XMM-Newton}, the remarkably deep net \textit{Chandra}/ACIS exposures in the immediate vicinity of NGC\,1404 (656.8 ks) offers the unique opportunity to compare the best-fitting properties obtained with each instrument within a common spectral region, hence to cross-check the consistency of our measurements. For each ACIS observation allowing so (see Table~\ref{tab:observations}), we thus extract a spectrum corresponding to the same core region. Unlike RGS, which integrates photons over the entire field-of-view diameter along its dispersion direction, we choose to restrict the EPIC and ACIS core regions to the typical extent of the galaxy (i.e. 2 arcmin), in order to minimise the counts-over-background ratio on these CCD observations. To avoid further technical complication in the spectral analysis, we also restrict our core EPIC analysis to the two deepest \textit{XMM-Newton} pointings only. Their dominant exposures (Table~\ref{tab:observations}) make us confident that this has negligible effect on our final estimates (i.e. less than $\sim$15\% improvement on our statistics), without adding much information given the predominance of systematic over statistical uncertainties for deep pointings.

The ``profiles'' regions consist of 10 EPIC concentric annuli centred on the galaxy's X-ray emission peak and covering 4~arcmin of total radius. These annuli are defined with arbitrary outer limits at 16, 31, 46, 61, 80, 105, 140, 175, 205, and 240~arcsec, as the result of a good compromise between (i) reasonably high S/N ratio in each annulus ($\sim$230 in external regions, up to $\sim$380 in the core) -- translating into relative $\Delta \mathrm{Fe}/\mathrm{Fe}$ uncertainties of 8--$14\%$, and (ii) delimiting regions of interest -- in particular the merging cold front at 105~arcsec ($\sim$10~kpc) from the core.

The 2D map is divided into 50 spatially independent EPIC cells, each having a (MOS+pn) constant S/N ratio of 200. The size and shape of the cells are obtained via the Weighted Voronoi Tesselation (WVT) binning method of \citet{diehl2006}. We ensure that (i) the smallest cells ($\sim$~14~arcsec of diameter) remain larger than the typical PSF of the EPIC instruments ($\sim$6~arcsec full width at half-maximum), (ii) the cells do not significantly overlap with the cold front, and (iii) all identified EPIC point sources were previously discarded from the map. In total, the map covers the whole $\sim$5~arcmin area surrounding the galaxy.

In addition, we define Eastern and Western sectors that are extracted in both ACIS and EPIC instruments (white dashed lines in Fig.~\ref{fig:NGC1404_zoomed}). The reasons for analysing these additional two sectors and the motivation for their angle and extent are further developed in Sect.~\ref{sec:discussion}.

\section{Spectral analysis}\label{sec:analysis}

The spectral analysis of our EPIC, RGS, and ACIS selected regions is performed via the \texttt{SPEX} fitting package \citep{kaastra1996} using the up-to-date Atomic Code and Tables \citep[SPEXACT v3.06;][]{kaastra2020} and the ionization balance of \citet{urdampilleta2017}. Before performing the spectral fitting itself, we use the \texttt{SPEX} auxiliary tool \texttt{trafo} to convert the raw spectra, the RMFs and the ARFs from the traditional OGIP format (respectively \texttt{.pi}, \texttt{.rmf} and \texttt{.arf}) into \texttt{SPEX}-readable files. 

The X-ray emission of NGC\,1404 is modelled with successively one (1T), two (2T), and three (3T) collisionally ionized plasma components \texttt{cie} and one power law \texttt{po}, all first redshifted and absorbed (\texttt{red*[hot*[cie+...+po]]}) and set to a distance of 18.79 Mpc (Sect.~\ref{sec:intro}). In the 3T case, we fix one temperature component to $kT_\mathrm{ICM} = 1.5$~keV to model the emission from the surrounding ICM of the Fornax cluster. These choices are compared, justified, and discussed further in Sect.~\ref{sec:results}. The power law, with its photon index set to 1.6 \citep[e.g.][]{su2017b}, is aimed to reproduce the integrated X-ray emission from the galaxy's low-mass X-ray binaries (LMXB). Although we model this component in all our selected annuli, we further verify its negligible contribution outside of the galaxy core (Sect.~\ref{sec:profiles}) and, therefore, fix it to zero in corresponding outer map cells. The redshift value is fixed to $z = 0.00649$ \citep{graham1998}. The reference hydrogen column density of the absorbing \texttt{hot} model is $n_\mathrm{H} = 1.57\times 10^{20}$~cm$^{-2}$ \citep{willingale2013}, which we allow to vary along a fixed grid within 0.51--$2.51\times 10^{20}$~cm$^{-2}$ as slight but significant offsets from the reference values are sometimes reported \citep[e.g.][]{mernier2016a,deplaa2017}. The electron temperature of this same model is set to a negligible value of 0.5~eV in order to mimic absorption by a neutral gas, while its abundance parameters are all kept to the Solar value. 

Our choice of the energy band to be considered depends evidently on the instrument. In the case of \textit{XMM-Newton} EPIC, we chose 0.6--10 keV, while our \textit{Chandra} ACIS spectra are fitted within 0.6--7 keV. In particular, we take care of avoiding setting the lower energy limit to 0.5 keV, as its proximity with the oxygen K-edge leads to erratic fitting and biased best-fitting estimates. Last but not least, we fit our RGS observations within 8--27~\AA~(roughly corresponding to 0.46--1.55 keV). Except the RGS spectra which are re-binned with a factor of 5 to reach 1/3 of the spectral resolution, we use the optimal binning method of \citet{kaastra2016} to analyse the EPIC and ACIS spectra. All fits are performed with C-statistics \citep{kaastra2017}.

Unless mentioned otherwise, the free parameters of our fits are (i) the emission measure\footnote{In \texttt{SPEX}, the emission measure is defined as $Y = \int n_e n_p dV$, where $n_e$ and $n_p$ are respectively the electron and proton densities, integrated over the entire source emitting volume.} of each \texttt{cie} component ($Y_\mathrm{gas,low}$, $Y_\mathrm{gas,high}$, and $Y_\mathrm{gas,ICM}$ when relevant), (ii) the temperature of one or two \texttt{cie} components ($kT_\mathrm{1T}$ for the 1T case; $kT_\mathrm{low}$ and $kT_\mathrm{high}$ for both the 2T and 3T cases), (iii) the emission measure of the \texttt{po} component ($Y_\mathrm{LMXB}$), and (iv) specific abundance parameters (Mg, Si, S, and Fe for the EPIC and ACIS spectra; N, O, Ne, Mg, Fe, and Ni for the RGS spectra), assumed for each element to be the same in all \texttt{cie} components\footnote{This assumption is admittedly simplified, and may not reflect the specific enrichment level of each component. High-resolution spectroscopy offered by future microcalorimeter instruments will enable to alleviate this constraint. Meanwhile, the impact of this assumption on our present results is expected to be very limited given the radially uniform abundances found in regions of multitemperature gas (Sect.~\ref{sec:profiles}).}. In every case, all the other $Z \ge 6$ abundances are tied to that of Fe. We stress the importance of coupling the O abundance in the EPIC and ACIS spectra to that of Fe in order to avoid critical biases in the fit. In fact, while the O line is essentially unresolved at CCD resolution, the high statistics of its corresponding spectral region forces the fit to reproduce its shape tightly, at the cost of incorrect estimates of other key features (e.g. temperature structure and/or abundances in the Fe-L complex).

\subsection{RGS line broadening}

Due to the slit-less nature of the gratings, the RGS spectra are affected by the spatial extent of the surface brightness (SB) of the source along its dispersion direction. This is a well-known effect, which naturally results in a broadening of the lines in the RGS spectra of extended sources \citep[e.g.][]{mernier2015,pinto2015}. The wavelength shift $\Delta\lambda$ can be expressed as
\begin{equation}
 \label{eq:RGS_broadening}
\Delta\lambda = \frac{0.138}{m} \Delta\theta{\AA},
\end{equation}
where $m$ is the order and $\Delta\theta$ is the angular shift\footnote{See the \textit{XMM-Newton} Users Handbook.}. This shift is then convolved with the entire SB profile of the source at this given wavelength. We model this instrumental broadening via the \texttt{lpro} component in \texttt{SPEX}, which we apply to the \texttt{cie} components. This multiplicative model relies on the SB profile of the MOS detectors along the dispersion direction, which was initially obtained using the \texttt{SPEX} auxiliary task \texttt{rgsvprof}. Two relevant parameters of this model are the scale parameter \texttt{s} and the spectral offset \texttt{dlam}\footnote{See the \texttt{SPEX} reference manual.}, both of which are left free in our RGS fits.

\subsection{Background treatment approaches}

In extended sources, it is crucial to treat the background carefully, as it can significantly impact our measurements at larger radii. 

As the RGS and ACIS data are mainly used to study the core region of NGC\,1404, background effects from these two instruments are expected to be limited. Therefore, we follow individually their standard prescriptions by extracting the background from, respectively, (i) ACIS blank-sky observations processed and re-pointed using the \texttt{CIAO} task \texttt{blanksky}, and (ii) RGS background templates of CCD\,9 (via the \texttt{SAS} task \texttt{rgsproc}), where presumably no source count is expected. We then rescale and subtract these background templates from their corresponding spectra.

The EPIC background deserves more careful attention, as we use these instruments also to explore the outer parts of the source. As detailed in \citet{mernier2015}, we model five well-known background components in our EPIC spectra; namely the Local Hot Bubble (LHB), the Galactic Thermal Emission (GTE), the Unresolved Point Sources (UPS), the residual Soft Protons (SP), and the Hard Particle (HP) background. The LHB and GTE are modelled respectively with an unabsorbed and absorbed \texttt{cie} component ($kT_\mathrm{LHB} = 0.07$~keV and $kT_\mathrm{GTE} = 0.20$~keV). The UPS is modelled with a power law of index $\Gamma_\mathrm{UPS} = 1.41$ \citep{moretti2003,deluca2004}. While these (foreground or background) components have an astrophysical origin, the SP and HP background components have an instrumental origin and, therefore, should \textit{not} be folded by the ARF of the instruments. The SP component is modelled with a power law of \textit{a priori} undetermined index (as it can vary between individual pointings and/or instruments) while the HP component is modelled with a broken power law and a series of Gaussian lines with values listed in \citet{mernier2015}. The normalizations and other \textit{a priori} unkown parameters are first left free and determined over the entire field of view, then fixed and rescaled accordingly to the selected regions, taking the vignetting curves of each detector into account where appropriate.


\section{Results}\label{sec:results}


\subsection{Galaxy core}\label{sec:core}

Besides its deep statistics, the very core of NGC\,1404 has the advantage of being observed by (\textit{XMM-Newton} and \textit{Chandra}) CCD spectrometers \emph{and} the RGS instrument. This provides an interesting opportunity not only to derive key thermal and chemical properties with unprecedented accuracy, but also to \emph{compare} them between all these instruments (i.e. EPIC MOS, EPIC pn, RGS, and ACIS), with the aim to better understand their most recent cross-calibration (at least in this cool temperature regime).

\begin{figure}
        \centering
                \includegraphics[width=0.49\textwidth, trim={0cm 0cm 0cm 0cm},clip]{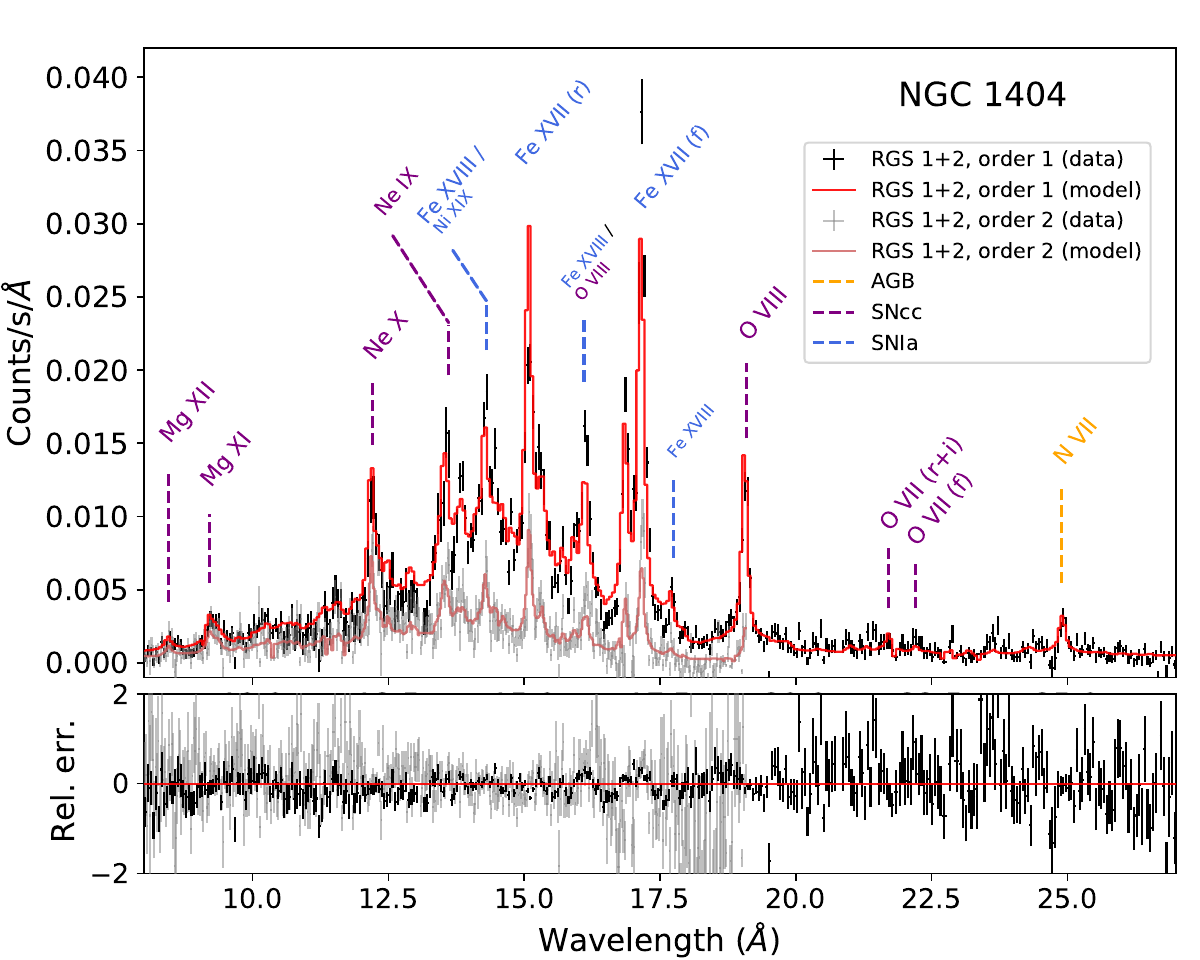}
        \caption{\textit{XMM-Newton} RGS fit of the core of NGC\,1404 (0.8~arcmin width, i.e. 90\% of the PSF). Both first and second orders are shown here. Prominent emission lines are labelled in orange, purple, or blue, depending on the stellar source of the concerned element (AGB, SNcc, or SNIa, respectively).}
\label{fig:RGScomb}
\end{figure}

\begin{figure}
        \centering
                \includegraphics[width=0.49\textwidth, trim={0cm 0cm 0cm 0cm},clip]{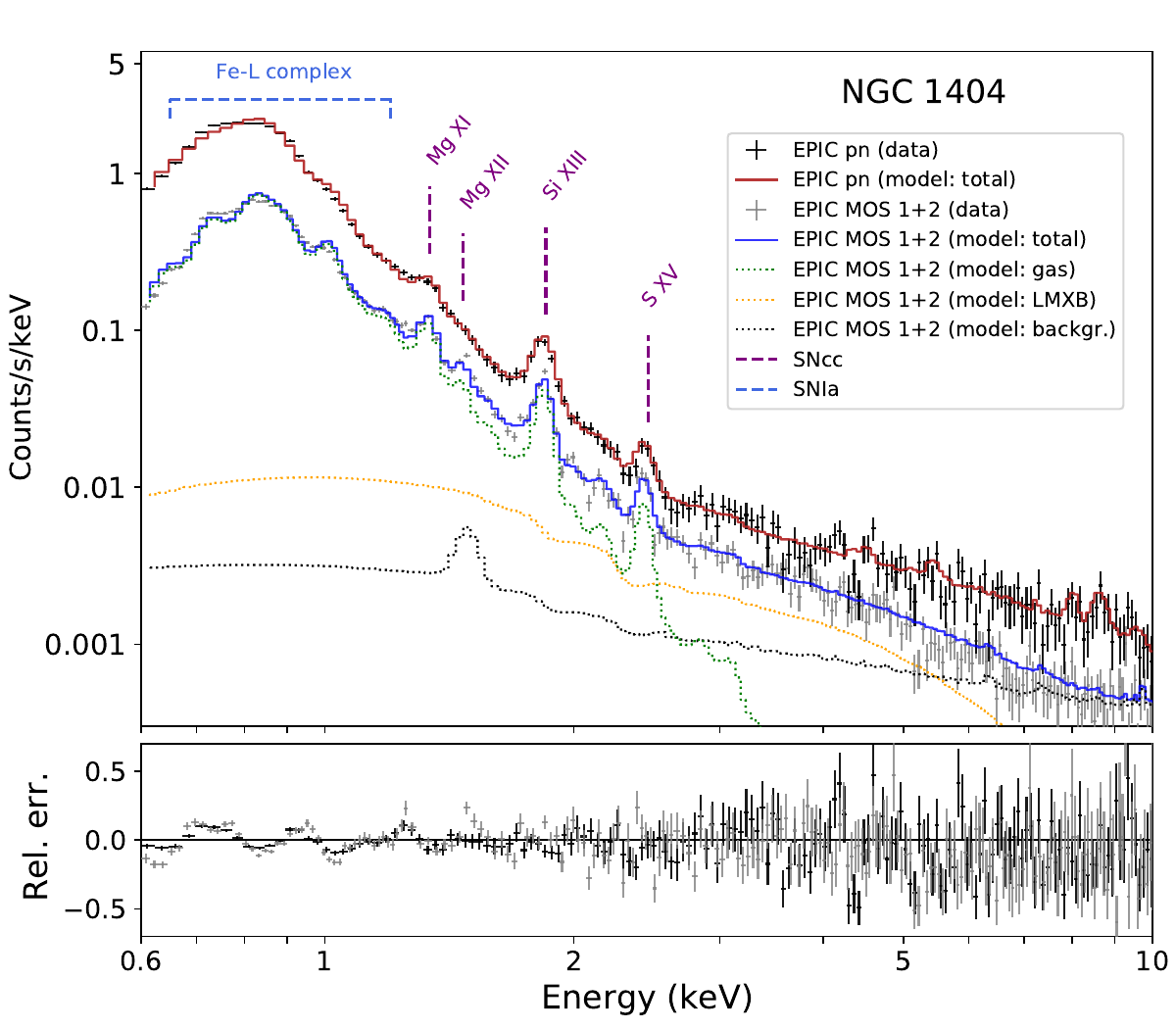}
        \caption{Combined \textit{XMM-Newton} EPIC fit of the core of NGC\,1404. The extraction region has been selected for the two deepest pointings to match that of RGS (0.8~arcmin width). Best-fitting models for the ICM emission (green), the LMXB component (yellow), and the (astrophysical and instrumental) background (black) are shown in dotted lines. For clarity, the spectral data of the MOS\,1 and MOS\,2 instruments have been stacked. Prominent emission lines/complexes are labelled in purple or blue, depending on the stellar source of the concerned element (SNcc, or SNIa, respectively).}
\label{fig:rgsbox_EPIC}
\end{figure}

The RGS order 1 and 2 spectra, shown in Fig.~\ref{fig:RGScomb} with a 3T modelling, exhibit a number of well-identified metal lines. We note, for instance, a clear signal at the expected N~VII line energy, allowing us to measure its abundance with decent accuracy. Similarly, and unlike some other ETGs, the Mg~XI and Mg~XII lines as well as the Ne~X line are prominent and well resolved. We note, however, the weak emissivity of the O~VII lines, suggesting a small amount of cooling gas below typically $kT \lesssim 0.5$~keV \citep[e.g.][]{pinto2014}. Despite the impressive resolving power of the instrument, its lower effective area translates into a less robust estimate of the continuum. While the \textit{absolute} abundances have thus limited constraints, the \textit{relative} X/Fe ratios do not directly depend on the equivalent width (hence on the continuum) of one line; therefore they can be measured accurately \citep[Table~\ref{tab:rgsbox}; see also e.g.][]{mao2021}. We also note from the RGS spectrum (Fig.~\ref{fig:RGScomb}) that the Fe XVII resonance line at 15~\AA\  is affected by a significant optical depth, resulting in resonant scattering \citep[see also][]{ogorzalek2017}, in turn leading to partial suppression of its line flux. After re-fitting the RGS spectrum excluding this line, however, we find that the changes on the Fe abundance and X/Fe ratios are marginal (i.e. less than $\sim$5\% and $\sim$10\%, respectively -- thus fully consistent with our statistical error bars).

\begin{figure*}
        \centering
                \includegraphics[width=0.85\textwidth, trim={2.1cm 0.6cm 2.7cm 1.3cm},clip]{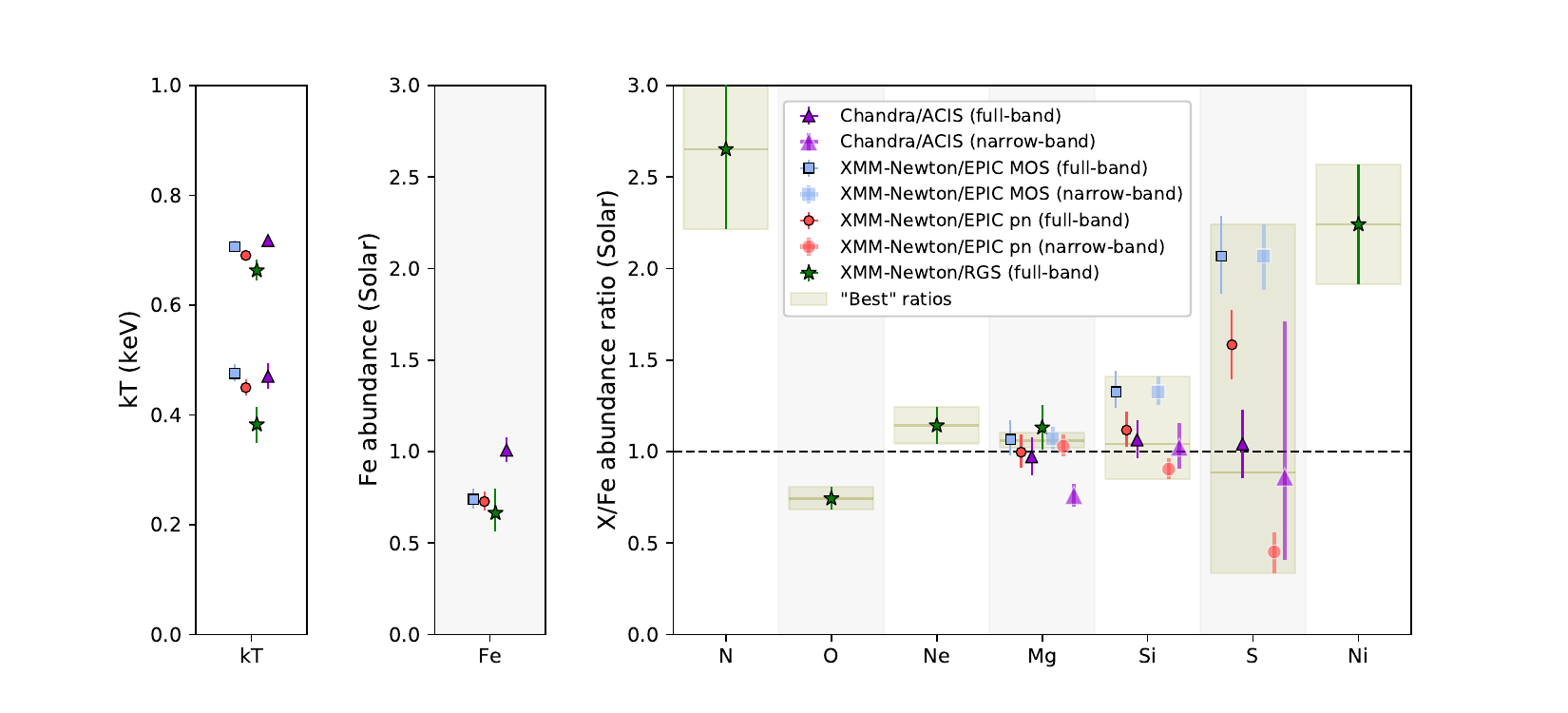}
        \caption{Best-fitting temperatures ($kT$), Fe abundances, and X/Fe abundance ratios obtained with \textit{XMM-Newton}/EPIC, RGS, and ACIS within a rectangular region covering the RGS extraction region of the deepest \textit{XMM-Newton} observation (see the text and Fig.~\ref{fig:NGC1404_zoomed}). The MOS, pn, and ACIS measurements are obtained over full-band fits for $kT$ and Fe and narrow-band fits for the abundance ratios (see the text).}
\label{fig:ratios}
\end{figure*}

\begin{table*}
\begin{centering}
\caption{Best-fitting results for the core region of NGC\,1404. The ACIS and EPIC extraction regions are chosen to coincide with the RGS extraction region (see the text and Fig.~\ref{fig:NGC1404_zoomed}). $^{f}$\,All abundances are estimated using full-band fits. $^{n}$\,Abundances other than Fe are estimated using narrow-band fits.}             
\label{tab:rgsbox}
\begin{tabular}{lccccc}        
\hline \hline                
Parameter & ACIS$^{n}$ & MOS$^{n}$ & pn$^{n}$ & EPIC (MOS+pn)$^{f}$ & RGS$^{f}$ \\    
\hline
$Y_\mathrm{gas,low}$ ($10^{69}$ m$^{-3}$)	&	$0.73 \pm 0.12$	&	$1.53 \pm 0.19$	&	$1.35 \pm 0.17$	&	$1.44 \pm 0.14$	&	$1.23 \pm 0.19$	\\
$Y_\mathrm{gas,high}$ ($10^{69}$ m$^{-3}$)		&	$1.64 \pm 0.12$	&	$2.05 \pm 0.23$	&	$2.4 \pm 0.3$	&	$2.17 \pm 0.18$	&	$2.29 \pm 0.18$	\\
$Y_\mathrm{gas,ICM}$ ($10^{69}$ m$^{-3}$)	&	$0.05 \pm 0.03$	&	$<0.016$	&	$<0.018$	&	$<0.009$	&	$<0.06$	\\
$Y_\mathrm{LMXB}$ ($10^{69}$ m$^{-3}$)	&	$0.492 \pm 0.021$	&	(EPIC)	&	(EPIC)	&	$0.638 \pm 0.017$	&	$2.14 \pm 0.21$	\\
$kT_\mathrm{low}$ (keV)				&	$0.484 \pm 0.02$	&	$0.476 \pm 0.016$	&	$0.450 \pm 0.017$	&	$0.462 \pm 0.012$	&	$0.38 \pm 0.03$	\\
$kT_\mathrm{high}$ (keV)					&	$0.723 \pm 0.010$	&	$0.706 \pm 0.012$	&	$0.690 \pm 0.011$	&	$0.699 \pm 0.008$	&	$0.664 \pm 0.021$	\\
Fe								&	$0.95 \pm 0.06$	&	$0.74 \pm 0.06$	&	$0.72 \pm 0.07$	&	$0.74 \pm 0.04$	&	$0.66 \pm 0.14$	\\
N/Fe								&	$-$	&	$-$	&	$-$	&	$-$	&	$2.7 \pm 0.5$	\\
O/Fe								&	$-$	&	$-$	&	$-$	&	$-$	&	$0.74 \pm 0.06$	\\
Ne/Fe							&	$-$	&	$-$	&	$-$	&	$-$	&	$1.14 \pm 0.10$	\\
Mg/Fe							&	$0.75 \pm 0.06$	&	$1.07 \pm 0.09$	&	$1.03 \pm 0.10$	&	$1.03 \pm 0.10$	&	$1.13 \pm 0.12$	\\
Si/Fe							&	$1.01 \pm 0.12$	&	$1.33 \pm 0.12$	&	$0.90 \pm 0.09$	&	$1.22 \pm 0.11$	&	$-$	\\
S/Fe								&	$0.9_{-0.5}^{+0.9}$	&	$2.0 \pm 0.3$	&	$0.43 \pm 0.16$	&	$1.87 \pm 0.20$	&	$-$	\\
Ni/Fe							&	$-$	&	$-$	&	$-$	&	$-$	&	$2.2 \pm 0.3$	\\
\hline                                   
\end{tabular}
\par\end{centering}
\end{table*}

The opposite situation takes place in the EPIC spectra. As illustrated in Fig.~\ref{fig:rgsbox_EPIC}, the moderate spectral resolution of the MOS\,1, MOS\,2 (hereafter, designated as ``MOS'' when fitted simultaneously) and pn instruments is compensated by their large spectral window, allowing a proper continuum estimate -- at least outside of the Fe-L complex. We also note that, whereas the flux of the hot gas emission dominates that of the other components by a factor of $>$20 in the 0.6--2 keV band, the LMXB component starts to dominate beyond $\sim$2 keV (with a 0.3--10~keV luminosity of $1.1 \times 10^{40}$~erg\,s$^{-1}$). The Si and, more importantly, the S abundance measurements are thus expected to be affected by the latter, if not perfectly modelled. The Mg abundances, however, are expected to be much more reliable. As we discuss further, the presence of fluorescent Al~K$\alpha$ instrumental lines in the EPIC spectra may seriously bias Mg measurements in the case of dominant (particle) background contributions; however the very high S/N ratio prevents this issue in such a central region (Fig.~\ref{fig:rgsbox_EPIC}). The same remarks apply to the ACIS spectra. 

\begin{figure}
        \centering
                \includegraphics[width=0.49\textwidth, trim={0.5cm 0cm 0cm 0.1cm},clip]{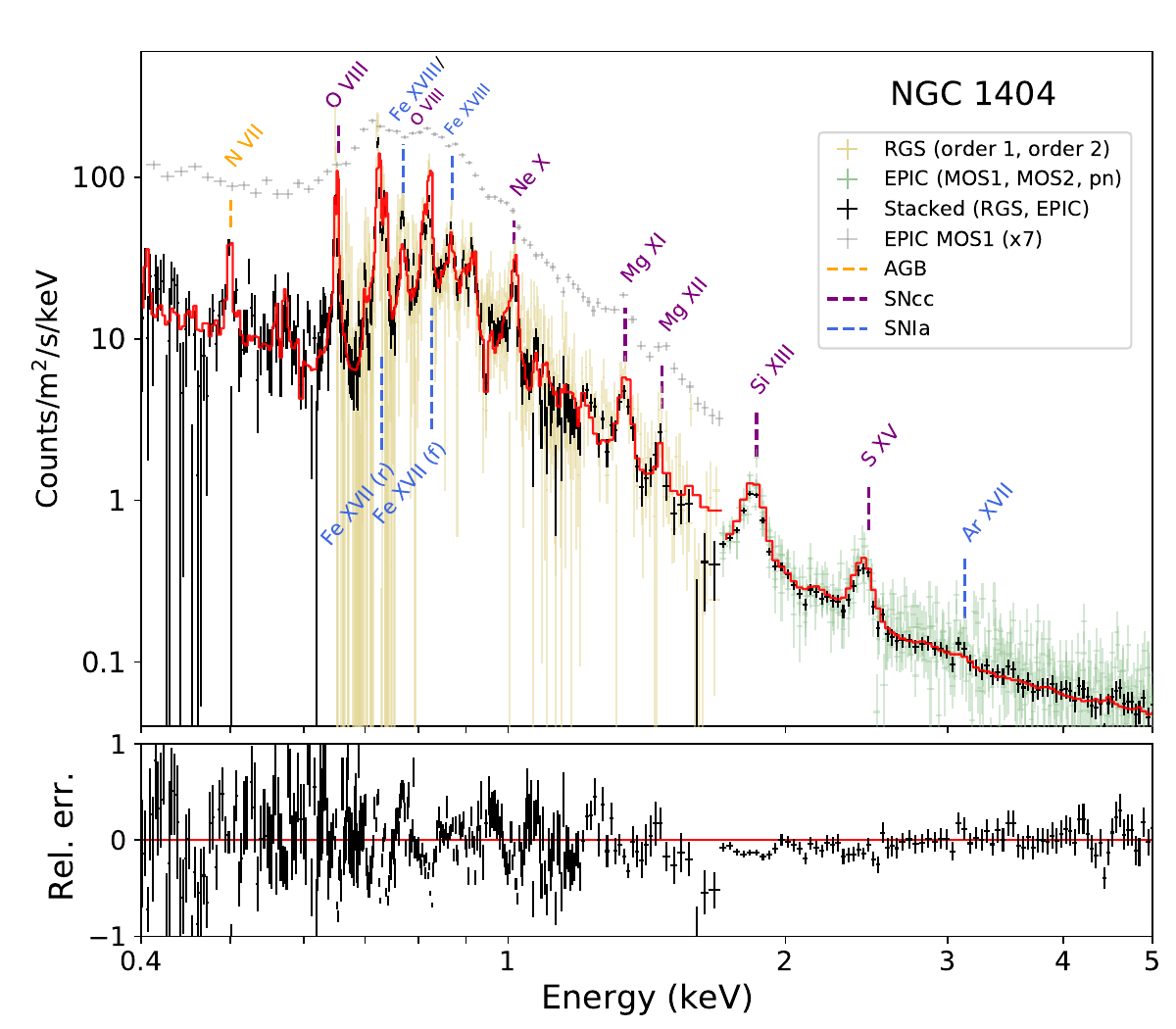}
        \caption{Combined \textit{XMM-Newton} RGS (1 \& 2) + EPIC (MOS\,1, MOS\,2, \& pn) fit of the core of NGC\,1404. The EPIC extraction region has been selected to match that of RGS (0.8~arcmin width). In order to keep all spatial regions consistent, only the deepest pointing has been used for RGS, while the two deepest pointings have been selected for EPIC (see the text). For clarity, the spectral data have been stacked over the different EPIC and RGS instruments. Line labels are the same as in Figs~\ref{fig:rgsbox_EPIC} and \ref{fig:RGScomb}. }
\label{fig:rgsbox}
\end{figure}

One common practice on measuring metal abundances (and particularly their X/Fe ratios) in high-quality CCD spectra is to fit the latter over the full spectral band. A more conservative way is to perform the same fit within a \textit{narrow} band, centred on the relevant emission line (i.e. 1--1.7~keV, 1.6--2.3~keV, 2.2--2.8~keV for Mg, Si, and S respectively). These ``narrow-band'' fits (in which the only free parameters are the relevant abundance and the total emission measure of the \texttt{cie} models\footnote{Throughout this study, narrow-band fits are performed after coupling $Y_\mathrm{gas,high}$ to $Y_\mathrm{gas,low}$, using the normalisation derived from its corresponding broad-band fit. This way, we ensure that the total emission measure is fitted locally with one parameter only, avoiding fitting degeneracies.}) have the advantage of accounting for subtle (yet non-negligible) miscalibration of the local effective area, which may have biased the estimate of the local continuum (hence of a few abundances) in the full-band fit \citep[e.g.][]{mernier2015,mernier2016a,simionescu2019a}. For evident reasons, narrow-band fits can be performed on well-resolved lines only, and are thus not possible on Fe (which relies on the Fe-L complex at this temperature regime).

Figure~\ref{fig:ratios} and Table~\ref{tab:rgsbox} compile and summarize the relevant best-fitting parameters and X/Fe abundance ratios obtained in this central region in the 3T case using these various instruments. These results are virtually identical to the 2T case, as $Y_\mathrm{gas,ICM}$ is found to be negligible here. For comparison, results from both full-band and narrow-band fits are displayed in Fig.~\ref{fig:ratios}. On the other hand, in Table~\ref{tab:rgsbox} we focus on ratios measured conservatively with narrow-band fits in each individual CCD instrument. Encouragingly, we find very similar (higher and lower) temperatures between all the instruments, with disagreements never exceeding $\sim$0.1 keV. The three \textit{XMM-Newton} instruments also agree remarkably well (within 1$\sigma$) in terms of absolute Fe abundance, which is measured around 0.74 Solar. A slight but significant difference appears when considering ACIS, the latter measuring Fe almost 30\% higher than the RGS and EPIC measurements. This effect is related to the best-fitting $Y_\mathrm{gas,high}$ and $Y_\mathrm{gas,low}$ parameters which are measured somewhat lower in ACIS than in EPIC (Table~\ref{tab:rgsbox}).

On the other hand, the X/Fe ratios provide somewhat contrasted results. First, we note the excellent agreement of Mg/Fe -- around 1~Solar -- between \textit{all} the \textit{XMM-Newton} measurements (regardless of the instrument and of the fitting methodology assumed here). The ACIS narrow-band Mg/Fe ratio is measured somewhat lower (around 0.75 Solar); however we recover the agreement when the ACIS absolute Mg abundance is normalized over the preferred EPIC/RGS Fe abundance. This further suggests that ACIS tends to overestimate the absolute Fe abundance, while leaving the (absolute) Mg abundance properly measured. Secondly, we note a larger scatter in our Si/Fe measurements, though formally consistent with being Solar. The situation deteriorates when considering the S/Fe measurements, spanning between $\sim$0.3 and $\sim$2.3 Solar. The discrepancies found in these latter two ratios are not surprising, as the Si-K and S-K equivalent widths are likely to be significantly affected by the background and/or the LMXB component(s) (see above). While this clearly affects the reliability of the S abundance, we choose to consider Mg and Si (as well as their respective X/Fe ratios) in the rest of the analysis. Finally, we report the N/Fe, O/Fe, Ne/Fe, and Ni/Fe ratios measured with RGS only. These ratios, along with the ones mentioned above, are further discussed in Sect.~\ref{sec:discu_composition}.

Last but not least, we also aim to fit the RGS and the EPIC spectra all \textit{simultaneously}, as shown in Fig.~\ref{fig:rgsbox}. Although this attempt leads to somewhat unstable fits (with e.g. difficulties to constrain two temperatures at the same time, given the imperfect calibration between the RGS and EPIC energy limits), we note that the measured X/Fe ratios are all formally consistent with our previous measurements. Beyond these numbers, this combined fit and its resulting figure may provide a useful glimpse (albeit at lower spectral resolution) of the spectral structure that will be routinely observed at this temperature regime by the next generation of X-ray microcalorimeters onboard \textit{XRISM} \citep{xrism2020} and, on longer term, \textit{Athena} \citep{barret2018}.

\subsection{Radial profiles}\label{sec:profiles}

\begin{figure*}
        \centering
                \includegraphics[width=0.49\textwidth, trim={0.cm 0.4cm 0.cm 0.3cm},clip]{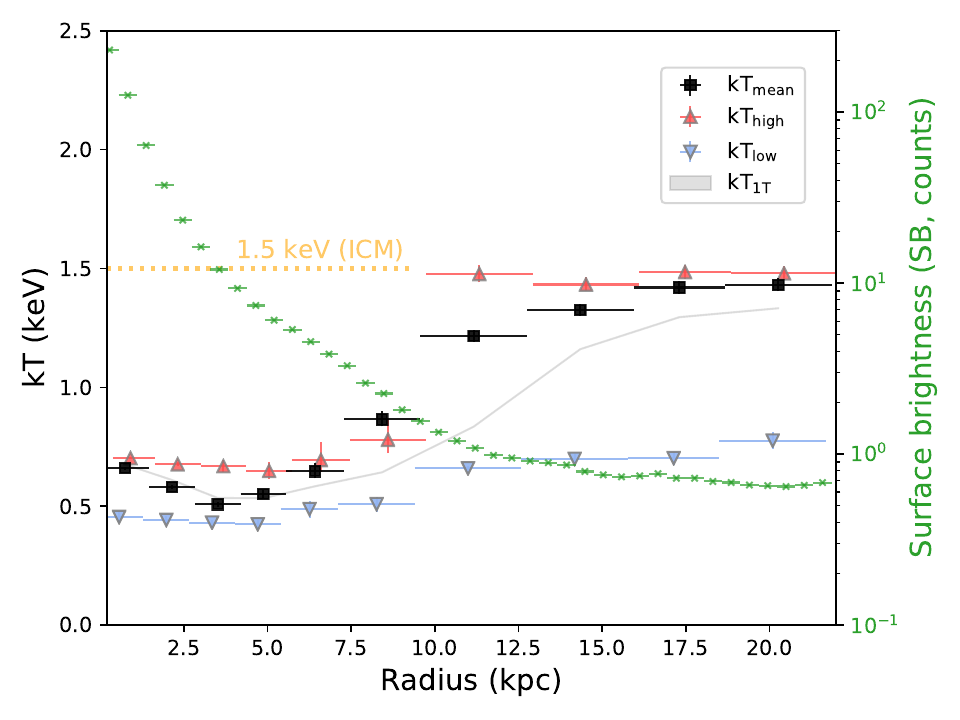}
                \includegraphics[width=0.49\textwidth, trim={0.cm 0.4cm 0.cm 0.3cm},clip]{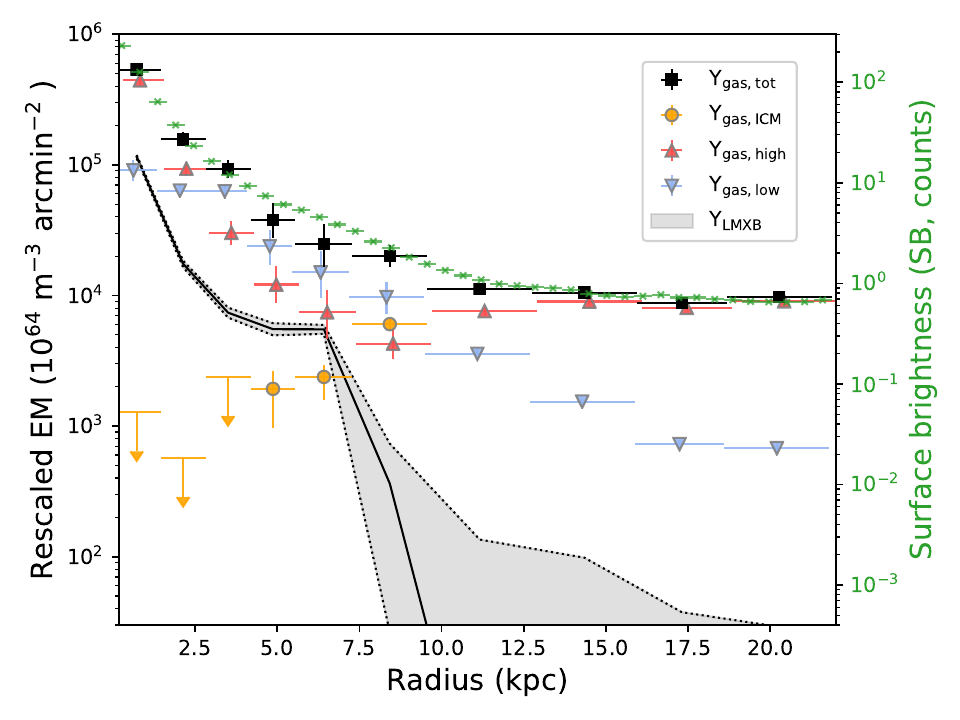} \\
                \includegraphics[width=0.49\textwidth, trim={0.cm 0.4cm 0.cm 0.3cm},clip]{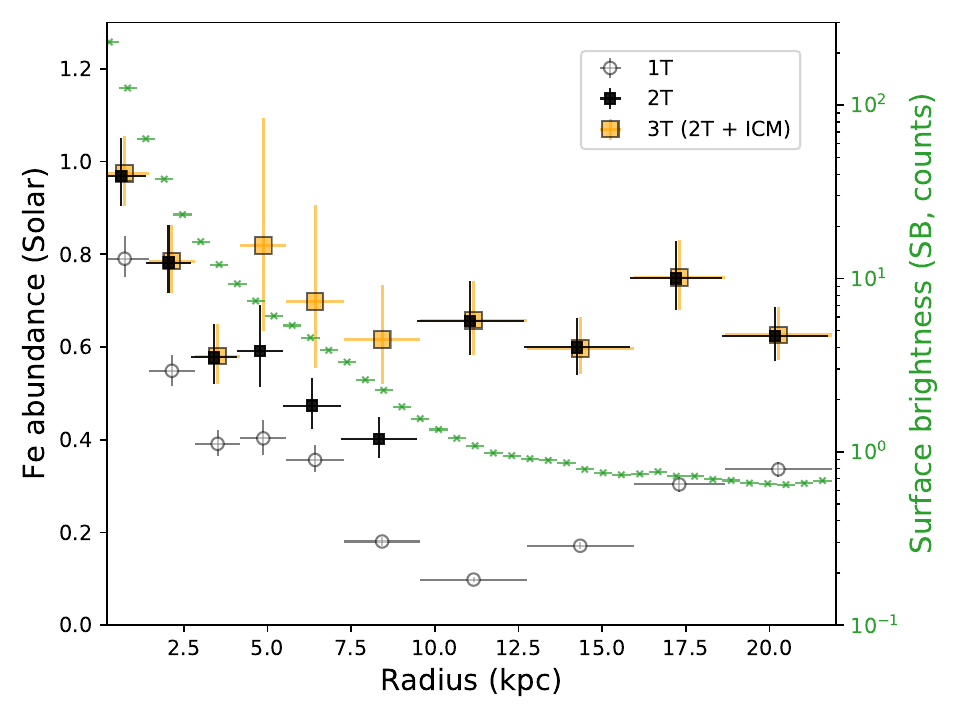}
                \includegraphics[width=0.49\textwidth, trim={0.cm 0.4cm 0.cm 0.3cm},clip]{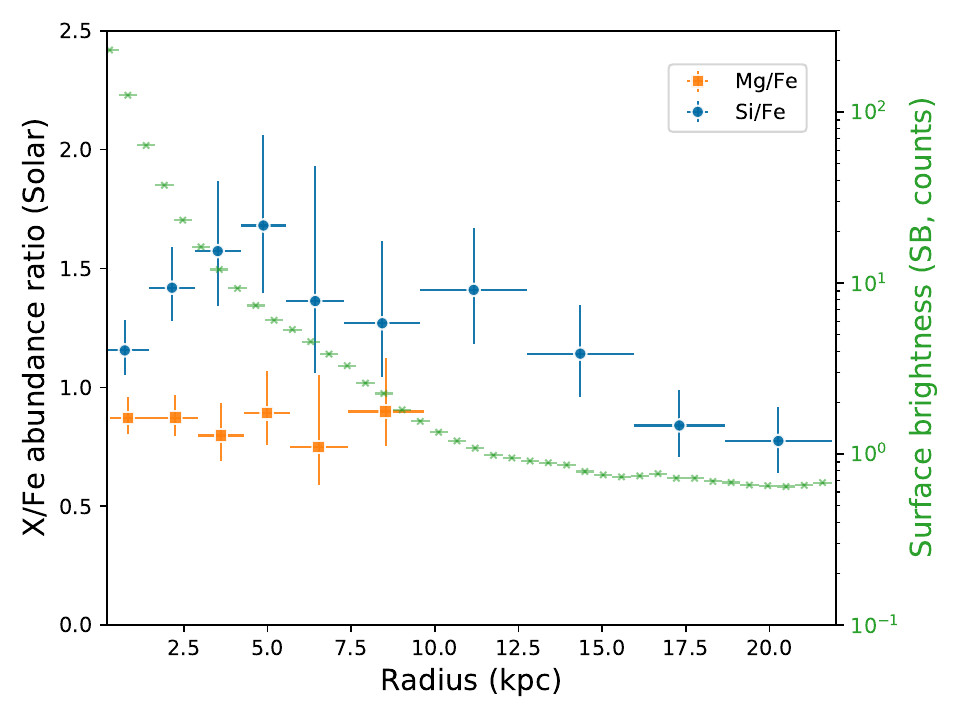}
        \caption{Radial profiles of NGC\,1404, analysed with \textit{XMM-Newton}/EPIC. In each plot, we also show the EPIC SB profile (green). Data points are slightly shifted along the x-axis for clarity. \textit{Top left-hand panel:} Temperatures of the free gas components, and their weighted mean. The temperature of the surrounding ICM component, fixed to 1.5~keV, is also shown for clairty. \textit{Top right-hand panel:} Emission measures (EM) for the different components (three hot gas components, one LMXB component), rescaled to their extracted area. The fixed value of the third thermal component, corresponding to the surrounding ICM, is also indicated (orange). \textit{Bottom left-hand panel:} Fe abundance. Results for 1T, 2T, and 3T fits are shown separately and discussed in the text. \textit{Bottom right-hand panel:} Mg/Fe and Si/Fe abundance ratios. The Mg and Si abundance parameters were fitted within a narrow energy band centred on their K-shell lines (see the text).}
\label{fig:profiles} 
\end{figure*}

The most relevant best-fitting parameters derived from our nine concentric annuli (Sect.~\ref{sec:analysis}) are shown in Table~\ref{tab:profiles} and Fig.~\ref{fig:profiles}. In the four panels of the figure, we also show (in green) the EPIC (MOS+pn) radial SB profiles.

\begin{table*}
\begin{centering}
\caption{Best-fitting results for our (azimuthally averaged) radial profiles of NGC\,1404, using the EPIC instruments. All parameters are derived using full-band fits of the EPIC (MOS+pn) instruments, except the Mg/Fe and Si/Fe ratios which include the conservative limits of (individual) MOS and pn narrow-band fits.}             
\label{tab:profiles}
\begin{tabular}{lccccccccc}        
\hline \hline                
Region & $Y_\mathrm{gas,low}$ & $Y_\mathrm{gas,high}$ & $Y_\mathrm{gas,ICM}$ & $Y_\mathrm{LMXB}$ & $kT_\mathrm{low}$ & $kT_\mathrm{high}$ & Fe & Mg/Fe & Si/Fe \\
(kpc) & ($10^{67}$ m$^{-3}$) & ($10^{67}$ m$^{-3}$) & ($10^{67}$ m$^{-3}$) & ($10^{67}$ m$^{-3}$) & (keV) & (keV) &  &  & \\   
\hline
0--1.5		&	$20 \pm 4$	&	$99 \pm 9$	&	$<0.3$	&	$25.7 \pm 0.8$	&	$0.45 \pm 0.03$	&	$0.701 \pm 0.007$	&	$0.97 \pm 0.08$	&	$0.87 \pm 0.08$	&	$1.16 \pm 0.13$	\\
1.5--2.8		&	$58 \pm 7$	&	$39 \pm 5$	&	$<0.4$	&	$10.7 \pm 0.6$	&	$0.442 \pm 0.016$	&	$0.677 \pm 0.011$	&	$0.79 \pm 0.08$	&	$0.87 \pm 0.09$	&	$1.42 \pm 0.17$	\\
2.8--4.2		&	$63 \pm 8$	&	$30 \pm 7$	&	$<2.4$	&	$7.5 \pm 0.7$	&	$0.431 \pm 0.013$	&	$0.668 \pm 0.025$	&	$0.58 \pm 0.07$	&	$0.80 \pm 0.13$	&	$1.6 \pm 0.3$	\\
4.2--5.6		&	$33 \pm 11$	&	$17 \pm 7$	&	$2.7 \pm 1.4$	&	$7.6 \pm 0.9$	&	$0.424 \pm 0.020$	&	$0.65 \pm 0.04$	&	$0.81_{-0.18}^{+0.28}$	&	$0.89_{-0.13}^{+0.18}$	&	$1.7 \pm 0.4$	\\
5.6--7.3		&	$35 \pm 16$	&	$17 \pm 8$	&	$5.6 \pm 1.8$	&	$12.9 \pm 1.0$	&	$0.49 \pm 0.04$	&	$0.69 \pm 0.08$	&	$0.70_{-0.14}^{+0.21}$	&	$0.75_{-0.16}^{+0.30}$	&	$1.4_{-0.3}^{+0.6}$	\\
7.3--9.6		&	$39 \pm 12$	&	$17 \pm 5$	&	$24.4 \pm 2.3$	&	$<3$	&	$0.51 \pm 0.03$	&	$0.78 \pm 0.09$	&	$0.62 \pm 0.12$	&	$0.89_{-0.15}^{+0.23}$	&	$1.27_{-0.22}^{+0.35}$	\\
9.6--12.8		&	$27 \pm 3$	&	$57 \pm 3$	&	$-$	&	$<1.0$	&	$0.659 \pm 0.009$	&	$1.48 \pm 0.04$	&	$0.65 \pm 0.09$	&	$-$	&	$1.4 \pm 0.3$	\\
12.8--15.9	&	$14.8 \pm 1.0$	&	$87 \pm 5$	&	$-$	&	$<0.9$	&	$0.699 \pm 0.020$	&	$1.43 \pm 0.03$	&	$0.60 \pm 0.06$	&	$-$	&	$1.14 \pm 0.21$	\\
15.9--18.7	&	$7.2 \pm 0.5$	&	$80 \pm 4$	&	$-$	&	$<0.4$	&	$0.70 \pm 0.03$	&	$1.48 \pm 0.03$	&	$0.75 \pm 0.08$	&	$-$	&	$0.84 \pm 0.15$	\\
18.7--21.8	&	$9.2 \pm 1.0$	&	$123 \pm 6$	&	$-$	&	$<0.4$	&	$0.78 \pm 0.04$	&	$1.48 \pm 0.03$	&	$0.62 \pm 0.06$	&	$-$	&	$0.77 \pm 0.14$	\\
\hline                                   
\end{tabular}
\par\end{centering}
\end{table*}

The top left-hand panel shows the radial distribution of the temperature of each component, as well as the ``mean'' temperature, defined as
\begin{equation}
kT_\mathrm{mean} = \frac{Y_\mathrm{gas,high} kT_\mathrm{high} + Y_\mathrm{gas,low} kT_\mathrm{low} + Y_\mathrm{gas,ICM} (1.5~\mathrm{keV}) }{Y_\mathrm{gas,high} + Y_\mathrm{gas,low} + Y_\mathrm{gas,ICM}}.
\end{equation}
While the cooler temperature shows a rather flat profile, the higher temperature jumps abruptly to 1.5~keV beyond $\sim$10~kpc. This 1.5~keV plateau corresponds to the temperature of the surrounding ICM, hence justifying \textit{a posteriori} our choice of fixing $kT_\mathrm{ICM}$ to this temperature in this work. Because $kT_\mathrm{high}$ converges to 1.5~keV in these outer annuli even when applying 3T fits, the virtually similar values hence obtained for $kT_\mathrm{high}$ and $kT_\mathrm{ICM}$ imply fitting degeneracies that should be avoided. For this reason, we chose to model all our $>$10~kpc regions with 2T only, with no impact on the rest of our results. This abrupt temperature transition, also seen in $kT_\mathrm{mean}$, is actually expected. Indeed, its location coincides with the well-known merging cold front lying at the NW edge of the galaxy's hot atmosphere \citep[e.g.][]{machacek2005,su2017a,su2017b}. For comparison, we also plot the (unique) temperature $kT_{1T}$ obtained when performing 1T fits in each annulus. While the internal $kT_{1T}$ profile follows remarkably well that of $kT_\mathrm{mean}$, larger deviations are observed directly outside of the core, with the former being systematically underestimated compared to the latter. Such discrepancies, which have been extensively discussed in previous work \citep[e.g.][]{vikhlinin2006}, arise in the regions where the gas is the \textit{least} isothermal (i.e.  where $Y_\mathrm{gas,high}$ and $Y_\mathrm{gas,low}$ are comparable while $kT_\mathrm{high}$ and $kT_\mathrm{high}$ differ by a factor of $>$2).

The top right-hand panel shows the emission measure of each modelled component. While no 1.5~keV gas is detected below $\sim$4~kpc, we see the relative contribution of $Y_\mathrm{gas,ICM}$ steadily increasing towards the outer regions of the galaxy. Interestingly, we note that the relative predominance of a given (cool vs. hotter) thermal component inverts twice across the whole profile. In fact, while the cooler component exceeds the hotter component by a factor of $\gtrsim$2 within $\sim$4--10~kpc (with similar temperatures), the hotter components dominate by an order of magnitude beyond $\sim$13~kpc \textit{and} within the innermost $\sim$1.5~kpc. Whereas the gas remains thus close to isothermal in the very core and in the ICM surrounding the galaxy, the intermediate regions show clear evidence of multitemperature structure. As naturally expected, the total gas emission measure $Y_\mathrm{gas,tot}$, simply defined as the sum $Y_\mathrm{gas,low} + Y_\mathrm{gas,high} + Y_\mathrm{gas,ICM}$, follows well the SB profile when assuming a proper normalisation factor between the two quantities. In addition, we also plot the derived emission measure of the LMXB component, $Y_\mathrm{LMXB}$. Compared with $Y_\mathrm{gas,tot}$, this component drops faster within the first $\sim$3~kpc, then flattens out to $\sim$6.5~kpc, before vanishing outside of the galaxy's SB as expected. It is not clear whether the intermediate flattening of the LMXB component is real or related to subtle fitting artefacts (e.g. intermediate thermal components); however we note that this component systematically lies a factor of $\gtrsim$5 below the total hot gas component. This makes us confident that LMXBs affect our present results only weakly.

The bottom left-hand panel shows the Fe abundance as measured in our simultaneous MOS+pn fits successively in the 1T, 2T, and 3T cases. We have verified that, for each case, individual MOS and pn fits provide consistent results. The lower abundance (and different profile shape) measured with 1T compared to multitemperature modelling is expected: in fact, metallicities measured from CCD-resolution spectra are well known to be underestimated if the temperature structure is modelled too simplistically with one component only \citep[the so-called ``Fe bias''; see e.g.][]{buote1994,buote2000,mernier2018c,gastaldello2021}. It is interesting to note (and, in fact, remind) that this bias occurs even when one gas component dominates the other by one order of magnitude (see the top right-hand panel). Much more surprising, however, is to find a significant Fe difference between the 2T and 3T cases within $\sim$4--10~kpc. This suggests that, at these intermediate radii where the ICM component competes with the galaxy's atmosphere, modelling spectra with 2T is not realistic enough. This ``double Fe bias'' is important to report as it significantly alters the shape of the Fe profile (particularly the apparent presence of a jump at $\sim$10~kpc) and its subsequent physical interpretations. The 3T profile appears to be centrally peaked, radially decreasing from $\sim$1~Solar in the very core down to $\sim$0.6~Solar outside of the galaxy and in the surrounding ICM.  For consistency check, we verify that 2T and 3T fits provide virtually identical Fe abundances in the surrounding ICM.

Finally, the bottom right-hand panel shows the radial distribution of the Mg/Fe and Si/Fe ratios, using the narrow-band fits from MOS and pn individually. The MOS and pn results are then combined in each annulus. Whereas the Si abundance can be reliably measured out to large distances (despite a relative proximity with the Si K$\alpha$ line at $\sim$1.75~keV in the MOS instruments), the brightest instrumental line Al K$\alpha$ coincides in energy with the Mg-K lines (i.e. $\sim$1.5 keV) in both detectors, making outer Mg abundance measurements less reliable. In fact, outside the sixth annulus the source-to-background counts ratio in the 1.4--1.6~keV band drops below 1.5. For this reason, we choose to restrict our Mg/Fe measurements to the innermost $\sim$10 kpc. 
Although we note that the Mg/Fe ratios are on average somewhat lower than in the central box region (Sect.~\ref{sec:core}), their radial distribution through the galaxy extent is remarkably flat, in line with what is typically reported for more massive systems \citep[e.g.][]{mernier2017}. A more surprising result, however, is that the Si/Fe ratio appears much less uniform, with hints for an enhancement at intermediate radii and/or a central drop. This is further discussed in Sections~\ref{sec:maps}~and~\ref{sec:discu_distribution}.

\subsection{Metal (and thermal) maps} \label{sec:maps}

In addition to the (azimuthally averaged) profiles described in the previous section, 2D spatial variations of the thermal and chemical properties of interest in NGC\,1404 are investigated.

\begin{figure*}
        \centering
               \includegraphics[width=0.4\textwidth, trim={0.cm 0.cm 0.cm 0.cm},clip]{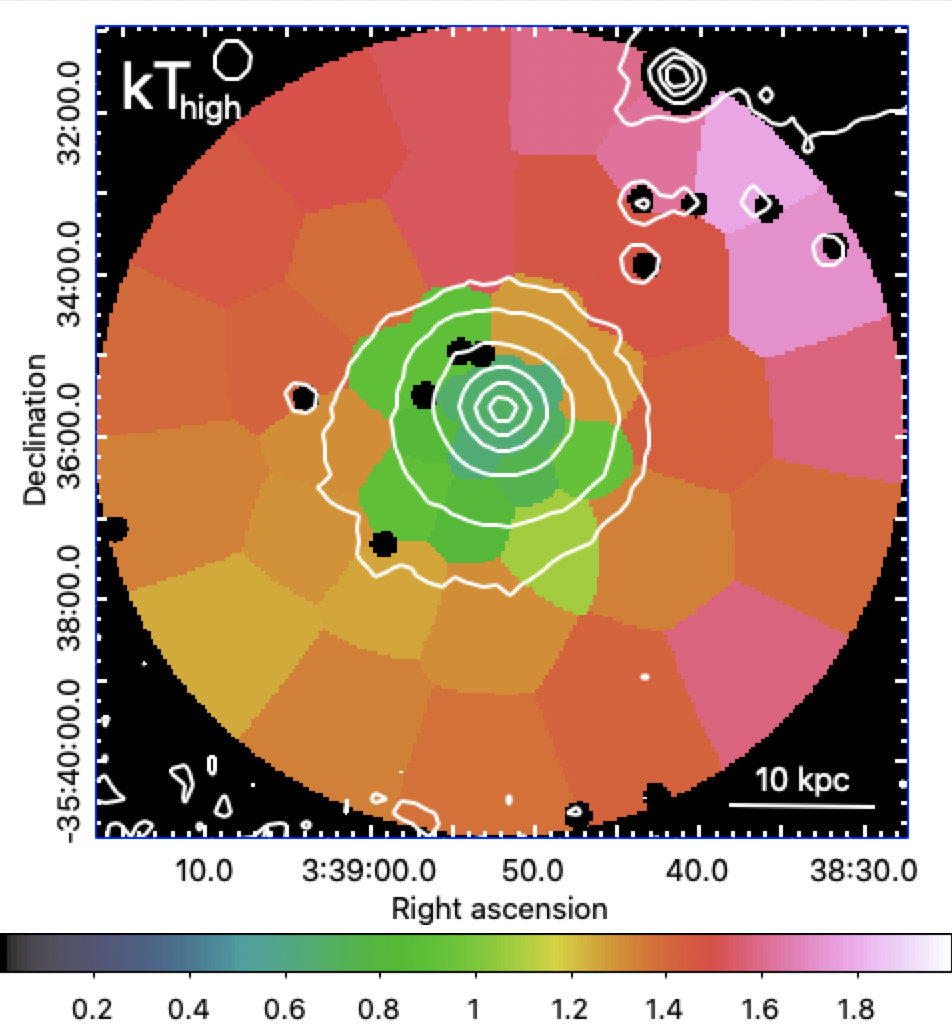}
                \includegraphics[width=0.4\textwidth, trim={0.cm 0.cm 0.cm 0.cm},clip]{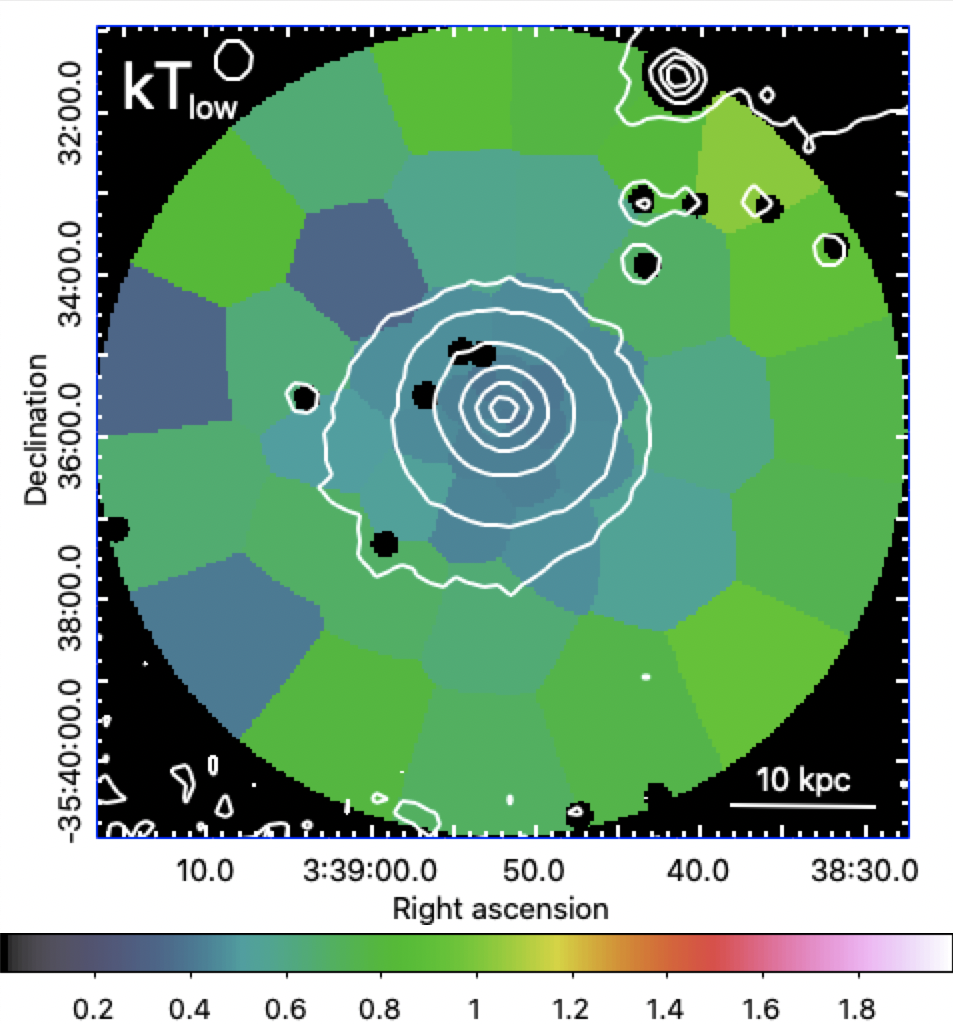} \\
                \includegraphics[width=0.4\textwidth, trim={0.cm 0.cm 0.cm 0.cm},clip]{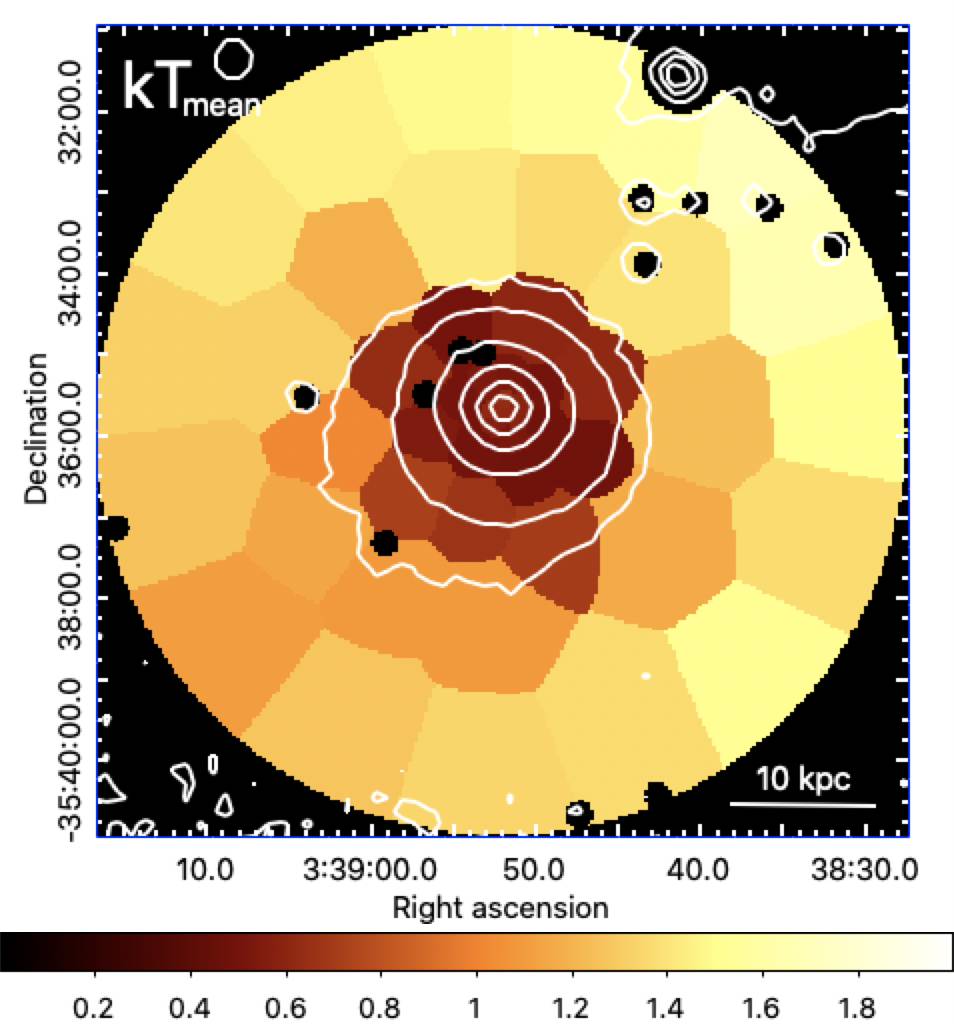}
                \includegraphics[width=0.4\textwidth, trim={0.cm 0.cm 0.cm 0.cm},clip]{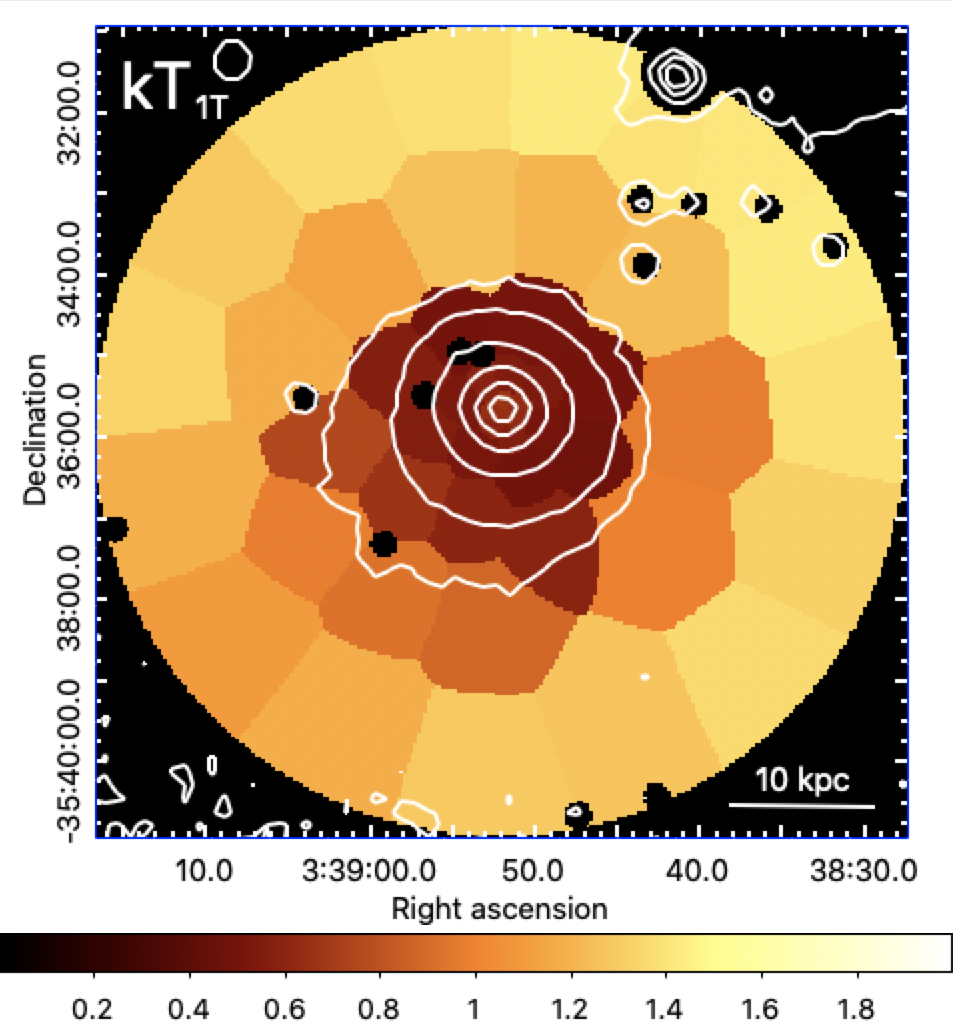}
        \caption{Temperature maps (in units of keV) obtained directly or indirectly from the EPIC fits (see the text). \textit{Top left-hand panel:} Higher temperature ($kT_\mathrm{high}$) measured with 3T fits (2T fits in the surrounding ICM). \textit{Top right-hand panel:} Lower temperature ($kT_\mathrm{low}$) for the same fits. \textit{Bottom left-hand panel:} Mean temperature ($kT_\mathrm{mean}$) measured with 3T fits (2T fits in the surrounding ICM). \textit{Bottom right-hand panel:} Temperature ($kT_\mathrm{1T}$) measured with 1T fits everywher on the map. The white contours are taken from the EPIC image and the cold front is approximately located on the NW edge of the outermost contour.}
\label{fig:mapsI}
\end{figure*}

Figure~\ref{fig:mapsI} focuses on the 2D thermal properties of the galaxy and its surroundings. In line with what is observed in our radial profiles, the higher temperature (top left-hand panel) shows important spatial variations between the ETG's hot gas and the surrounding ICM. Interestingly, we note that this component becomes somewhat cooler along the ram-pressure tail. The lower temperature (top right-hand panel), on the other hand, remains more uniform across the full map extent (though with a slight decrease toward the centre) -- again in agreement with our azimuthally averaged profiles. Combining these two temperatures (as well as $kT_\mathrm{ICM}$ in the inner $\sim$10~kpc cells), the mean temperature (bottom left-hand panel) reveals the dichotomy between the NGC\,1404's hot atmosphere and the Fornax ICM. In particular, we note little variation within the galaxy extent, while a clear temperature jump is seen at the interface of the NW merging cold front. The ram-pressure tail, somewhat cooler than its surroundings, is also clearly visible. Although comparable at first glance, this 3T map is not identical to its 1T counterpart (bottom right-hand panel). Similar to our radial analysis, we find that $kT_\mathrm{1T}$ is underestimated in intermediate regions and we suspect this bias to be due to its multitemperature structure.

\begin{figure*}
        \centering
                \includegraphics[width=0.4\textwidth, trim={0.cm 0.cm 0.cm 0.cm},clip]{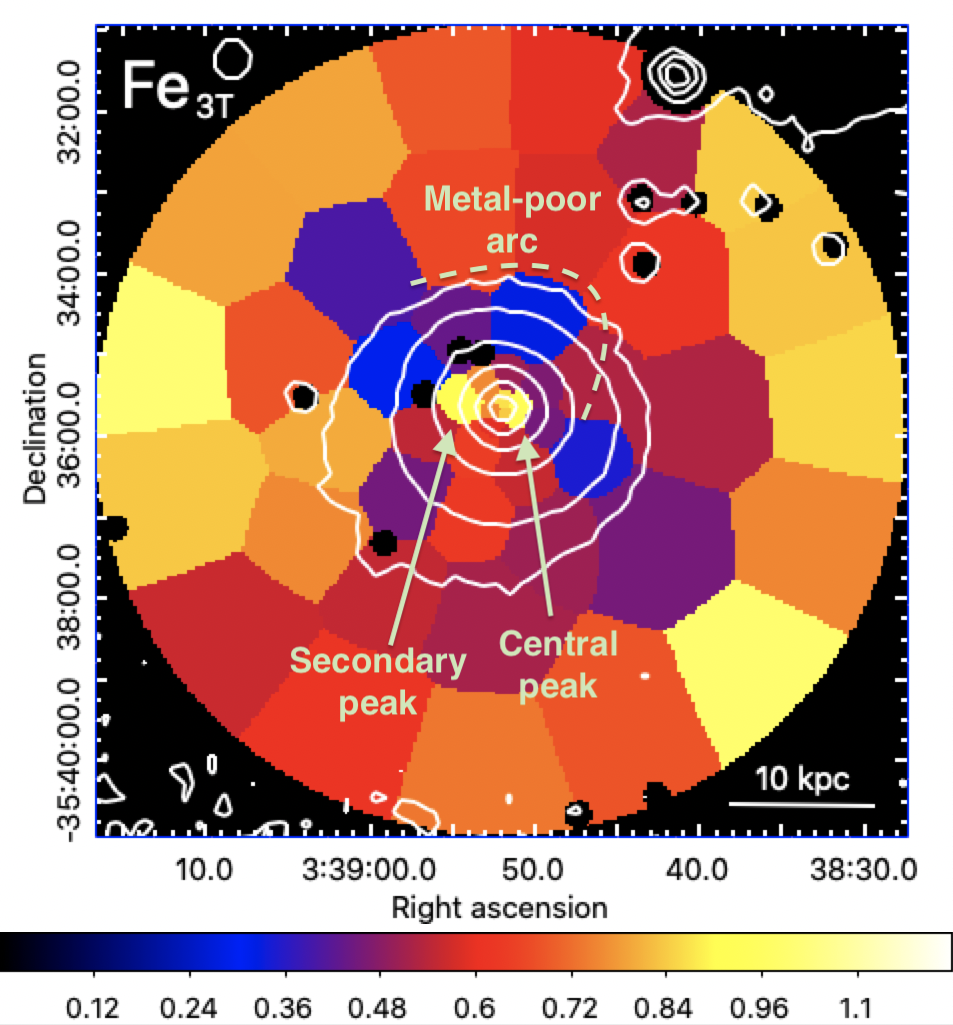}
                \includegraphics[width=0.4\textwidth, trim={0.cm 0.cm 0.cm 0.cm},clip]{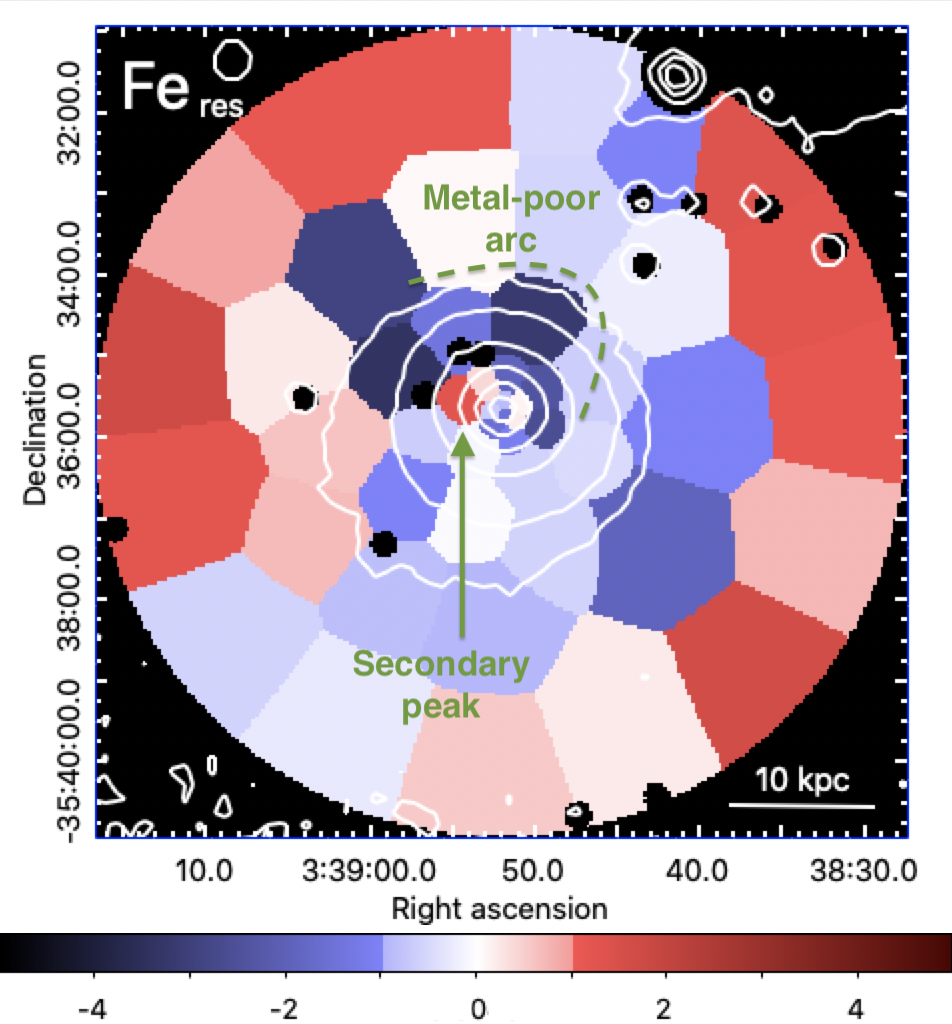} 
        \caption{Fe abundance map (in Solar units) derived with the EPIC instruments using either 3T fits (left-hand panel) and its residuals compared to the azimuthally averaged profile (right-hand panel -- see the text). The map has been colour-coded to emphasize cells with $>$1$\sigma$ residuals. Noteworthy features are annotated on the figures.}
\label{fig:mapsII}
\end{figure*}

Figure~\ref{fig:mapsII} focuses on the Fe abundance, obtained through our 3T (and partly 2T) fitting approach described above. The absolute Fe abundance map, shown in the left-hand panel, reveals a complex structure. Besides the rather large fluctuations seen in the surrounding ICM, we note the apparent presence of a secondary Fe peak $\sim$3~kpc East of the central (primary) X-ray peak. Last but not least, regions located directly under the NW cold front seem to exhibit systematically lower abundances (hereafter, the metal-poor arc). In order to better quantify the significance of these features, in the right-hand panel we derive the spatial Fe abundance \textit{residuals} with respect to their corresponding value expected from the azimuthally averaged radial profile. These residuals are expressed as 
\begin{equation}
\mathrm{Fe}_\mathrm{res} = \frac{\mathrm{Fe} - \mathrm{Fe}_\mathrm{exp}(r_\mathrm{cell})}{\Delta\mathrm{Fe}},
\end{equation}
where $\Delta\mathrm{Fe}$ is the statistical error on the measured Fe abundance and $\mathrm{Fe}_\mathrm{exp}(r_\mathrm{cell})$ is a functional form of the Fe radial profile (consisting of a central peak and a constant) fitted from our 3T Fe profile (Fig.~\ref{fig:profiles}, bottom left-hand panel), taken at the radial position of the centre of the considered cell, $r_\mathrm{cell}$. Defined this way, absolute values higher than 1 translate into $>$1$\sigma$ deviation from the expected azimuthally averaged profile. Interestingly, it appears that the three features mentioned earlier (i.e. overall ICM fluctuations, central and secondary peaks, and the metal-poor arc) are all significant. This is particularly true for the metal-poor arc, with cells measured $>$3$\sigma$ under the azimuthally averaged profile. We note that spectra in these cells have been fitted with 3T, hence should \textit{not} be sensitive to the double Fe bias reported above. This latter feature is further discussed in Sect.~\ref{sec:discu_distribution}.

\begin{figure*}
        \centering
               \includegraphics[width=0.4\textwidth, trim={0.cm 0.0cm 0.cm 0.cm},clip]{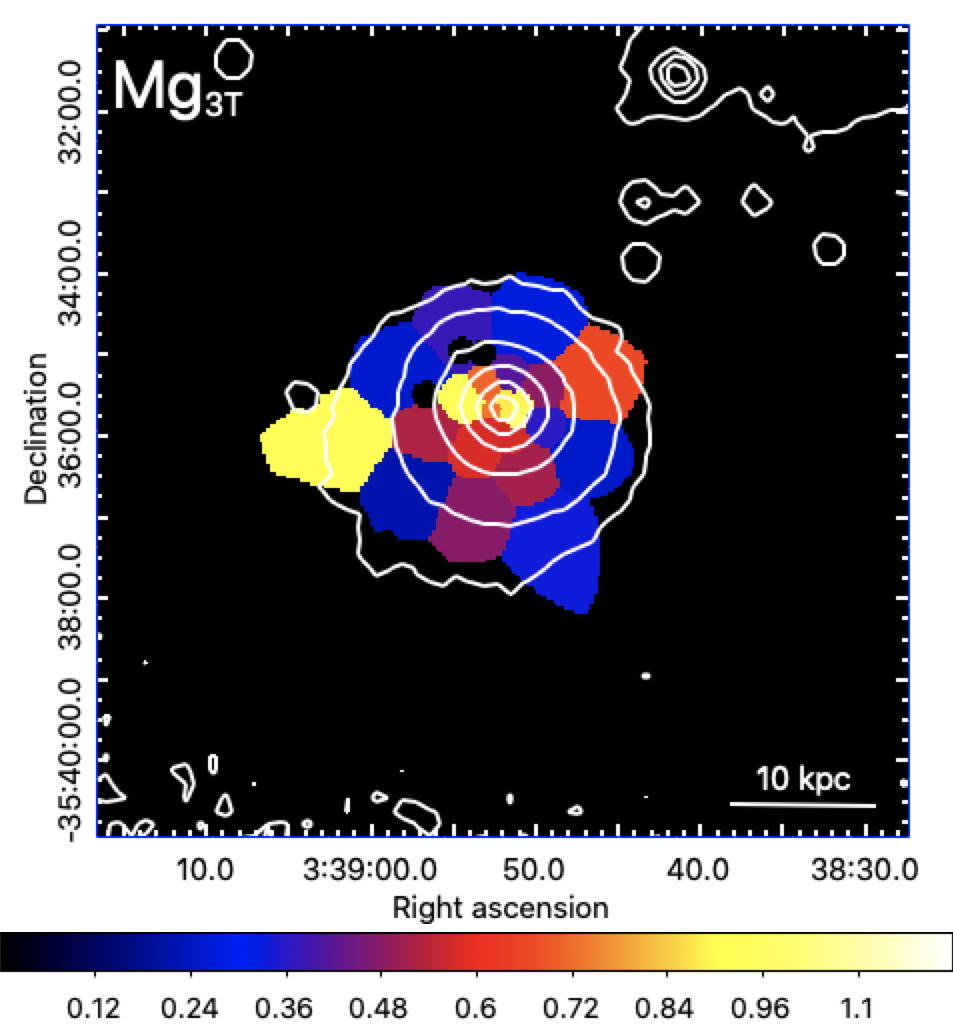}
                \includegraphics[width=0.4\textwidth, trim={0.cm 0.0cm 0.cm 0.cm},clip]{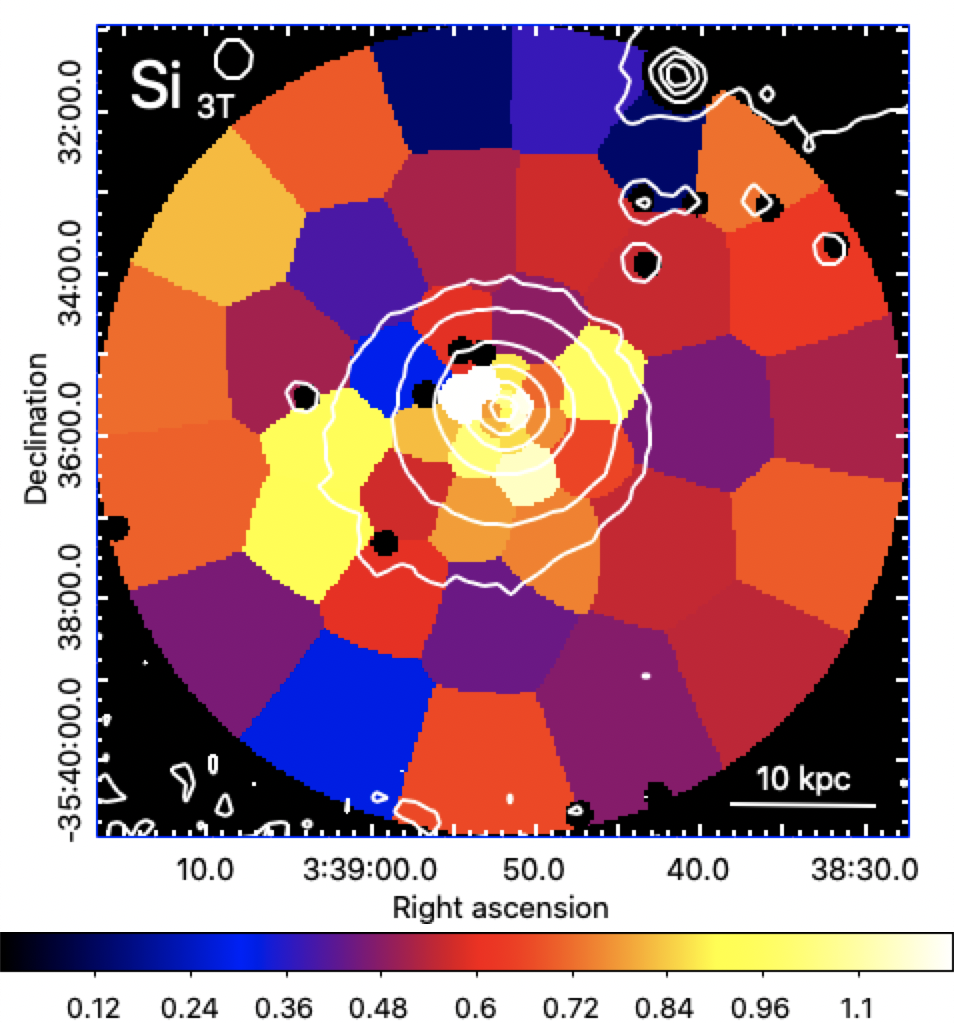} \\
                \includegraphics[width=0.4\textwidth, trim={0.cm 0.0cm 0.cm 0.cm},clip]{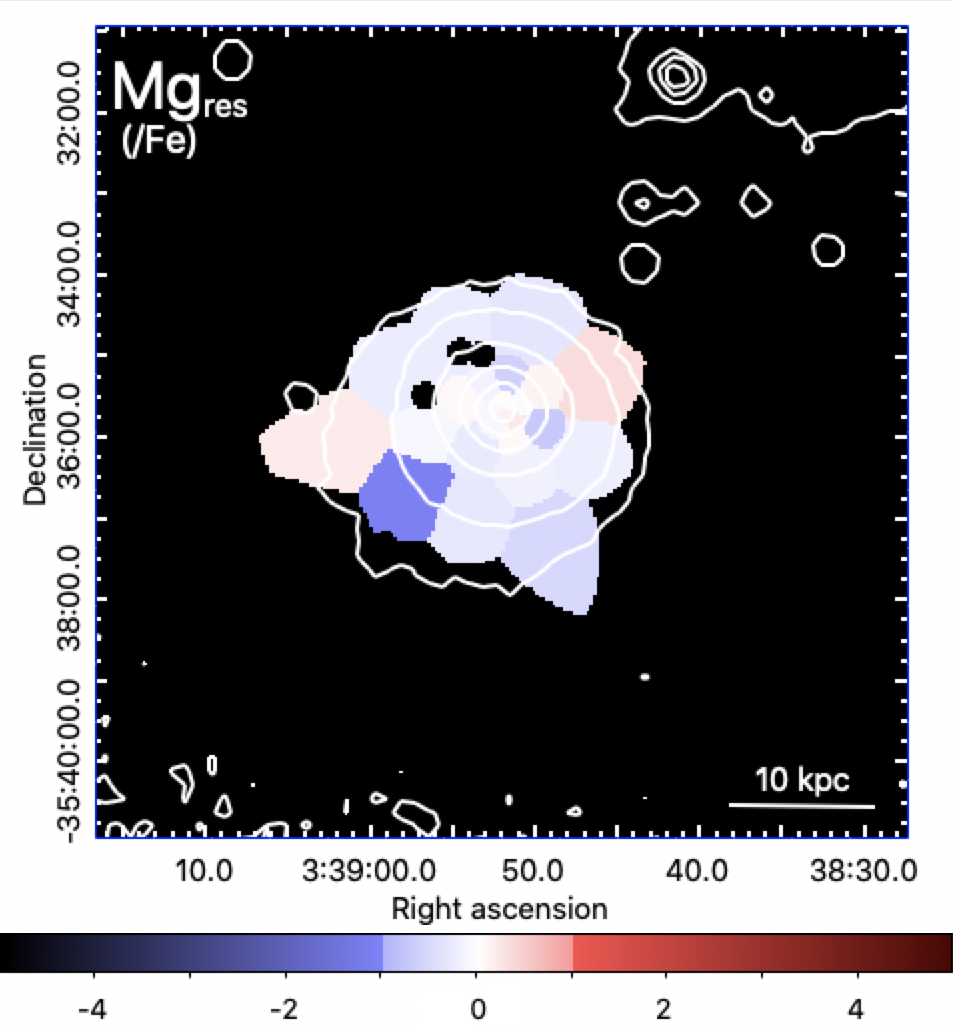}
                \includegraphics[width=0.4\textwidth, trim={0.cm 0.0cm 0.cm 0.cm},clip]{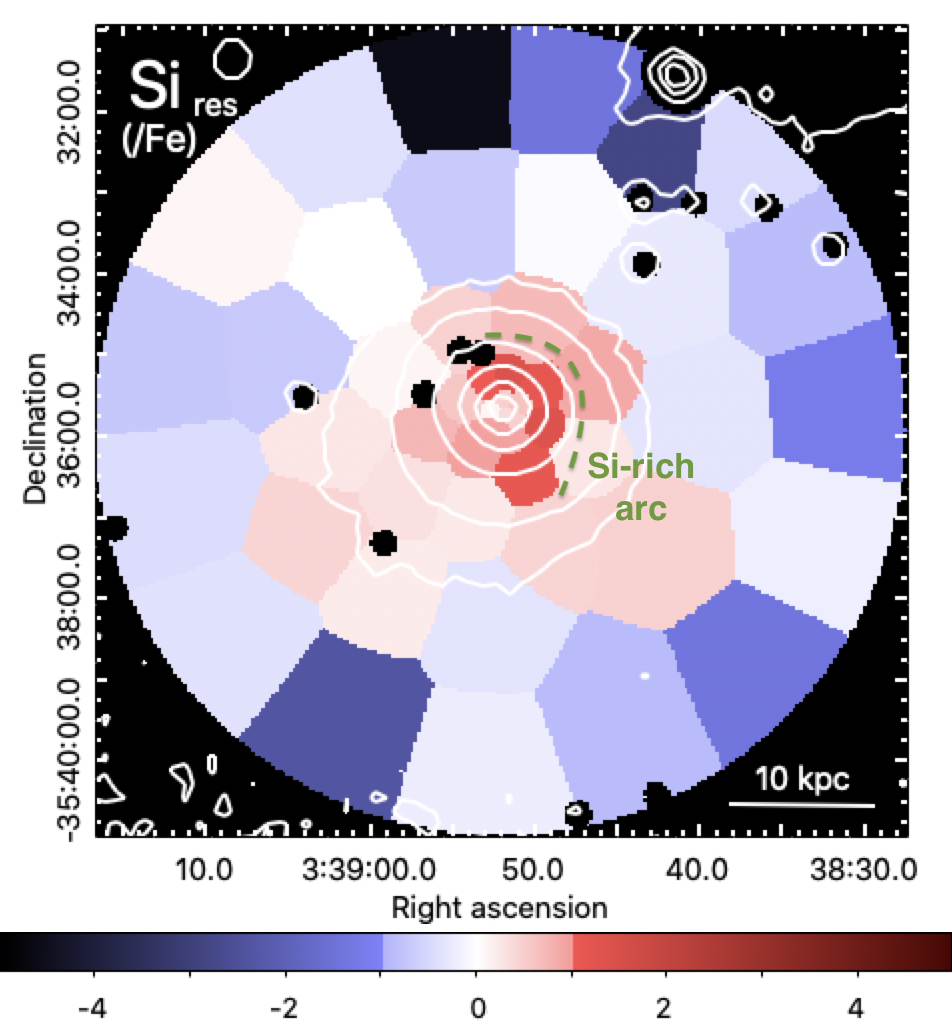}
        \caption{Abundance maps of Mg (\textit{top left-hand panel}) and Si (\textit{top right-hand panel}), and of the deviations (or residuals -- see the text) of their ratios $\mathrm{(X/Fe)}_\mathrm{res}$ (\textit{bottom left-hand panel}, \textit{bottom right-hand panel}). The Si-rich arc discussed in the text is annotated on the bottom right-hand panel.}
\label{fig:mapsIII}
\end{figure*}

Figure~\ref{fig:mapsIII} focuses on the Mg and Si maps (top left-hand and right-hand panels). In order to maximise the statistics, and given the relatively consistent measurements between our two fitting methodologies, these abundances are derived in their full-band fits. Like Sect.~\ref{sec:profiles}, we consider Mg only within the SB extent of NGC\,1404. Interestingly, the Mg map also indicates the central \textit{and} secondary peaks as mentioned above for Fe. However, these maps seem at first glance to show a slightly different pattern than the Fe map. To quantify this further, and similar to the Fe residuals described above, in the bottom panels of Fig.~\ref{fig:mapsIII} we express the spatial residuals of the Mg/Fe and Si/Fe ratios (left-hand and right-hand panels, respectively) with respect to the Solar threshold approached in the core (Fig.~\ref{fig:ratios}). These residuals are thus defined as 
\begin{equation}
\mathrm{(X/Fe)}_\mathrm{res} = \frac{\mathrm{(X/Fe)} - 1}{\Delta\mathrm{(X/Fe)}},
\end{equation}
where $\Delta\mathrm{X/Fe}$ is the statistical error on the measured X/Fe abundance ratio. 
The $\mathrm{(Mg/Fe)}_\mathrm{res}$ case shows that all investigated cells are almost all consistent with the Solar ratio, in line with the constant Mg/Fe profile seen in Fig.~\ref{fig:profiles} (bottom right-hand panel). Even though one cell exhibits slight >1$\sigma$ residuals, this number is statistically expected given the number of cells (24) within the galaxy extent. The $\mathrm{(Si/Fe)}_\mathrm{res}$ case, on the other hand, is more surprising. One can easily notice an extended arc-like region with significant $\mathrm{Si/Fe} > 1$ values. While this Si-rich arc seems to wrap around the very core, we note that the latter remains formally consistent with $\mathrm{Si/Fe} = 1$, as suggested by our core analysis (Sect.~\ref{sec:core}). This feature is, in fact, supported by the radial excess of Si/Fe reported in Sect.~\ref{sec:profiles} (Fig.~\ref{fig:profiles} bottom right-hand panel), and will be further investigated and discussed in Sect.~\ref{sec:discu_distribution}.


\section{Discussion}\label{sec:discussion}


\subsection{The peculiar metal distribution in and around NGC\,1404}\label{sec:discu_distribution}

\subsubsection{Central metal peak and ram-pressure stripping}\label{sec:central_peak}

\begin{figure*}
        \centering
                \includegraphics[width=0.49\textwidth, trim={0cm 0cm 0cm 0.1cm},clip]{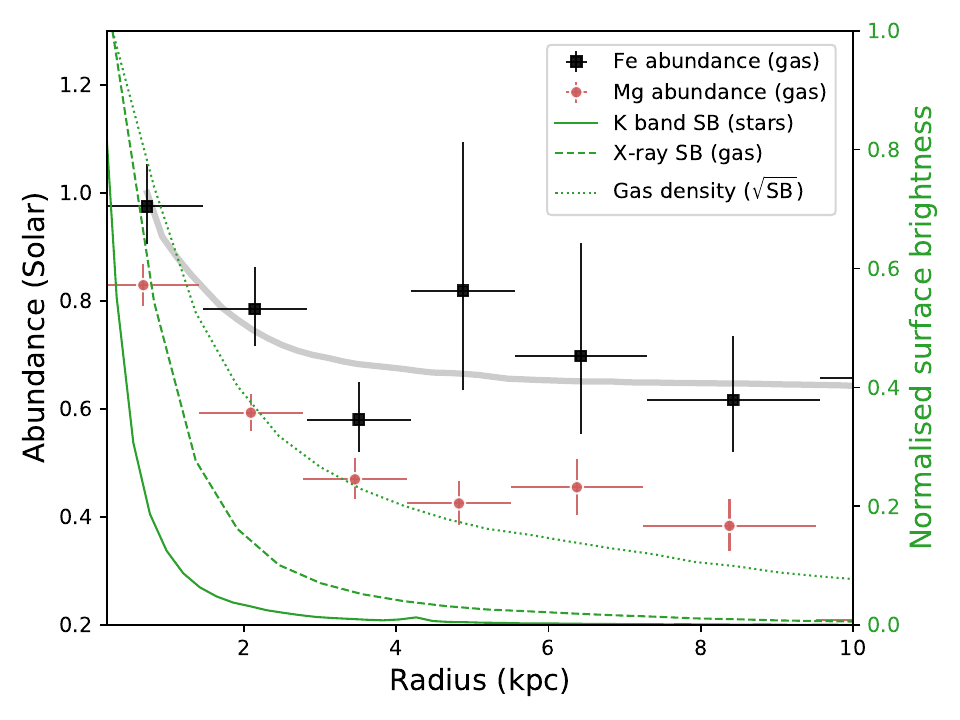}
                \includegraphics[width=0.49\textwidth, trim={0cm 0cm 0cm 0.1cm},clip]{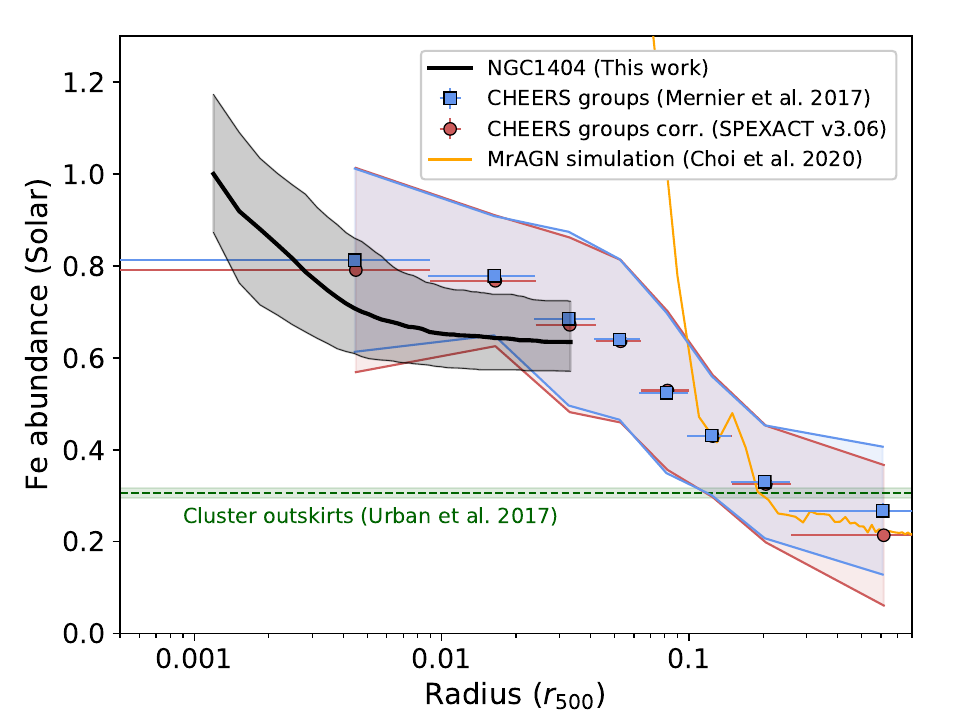}
        \caption{Left-hand panel: (Zoomed-in) Fe and Mg radial profiles of NGC\,1404, compared with the X-ray (gas) and K band (2MASS; stars) SB profiles. In order to be compared in a consistent way, these two SB profiles are normalized to their respective peak value. For consistent comparison, the (normalized) square-root of the X-ray SB, tracing the gas density, is also shown. The grey curve corresponds to the middle quartile of a series of fits over the range of uncertainties of the Fe radial profile (see the text). Right-hand panel: Full Fe radial profile (with is lower, middle, and upper estimated quartiles) compared with the average radial profile of 21 galaxy groups and ETGs \citep[CHEERS;][]{mernier2017}. For consistent comparison, this average profile is also corrected from the most recent SPEXACT version (v3.06, see the text). The average uniform metallicity in cluster outskirts is also shown for comparison \citep[][]{urban2017}, as well as recent simulation results from \citet{choi2020} including AGN feedback.}
\label{fig:profiles_discussion}
\end{figure*}

The Fe distribution is centrally peaked -- a feature that is commonly found in the hot gas of other (cool-core) galaxy groups and ETGs. This peak is better parametrized in Fig.~\ref{fig:profiles_discussion}, based on a series of Monte Carlo radial fits on the Fe profile, assuming possible values to be distributed within the 1$\sigma$ uncertainties of the Fe abundance measurements.

Zooming on the inner 10~kpc, (Fig.~\ref{fig:profiles_discussion} left-hand panel), we notice that the Fe distribution is comparable to that of the gas density, inferred from taking the normalized square-root of the X-ray SB\footnote{Although this comparison needs to be taken with caution (as, in addition ot the continuum, the SB also includes a non-negligible fraction of metal line emission), we remind the good correspondence between $Y_\mathrm{gas,tot}$ and the X-ray SB (Fig.~\ref{fig:profiles} top left).}. However, it is clearly broader than the stellar distribution, as traced by the K band luminosity of the galaxy (obtained publicly from the 2MASS survey), as well as its effective radius \citep[$R_e \simeq 2.3$~kpc;][]{sarzi2018}. Both these comparisons are also found in other systems, in particular relaxed clusters \citep[e.g.][]{degrandi2014}. In line with the uniform Mg/Fe ratio (Fig.~\ref{fig:profiles}, bottom right-hand panel), the radial Mg profile is found to have a shape remarkably similar to that of Fe.

Comparing our present Fe profile with the 21 groups/ETGs of the CHEERS sample, NGC\,1404 shows a rather unusual metal distribution. Indeed, although the surrounding ICM has its Fe abundance comparable to that seen in other systems at the same distance, the central Fe peak of NGC\,1404 seems narrower and flattens at small radii. This difference seems independent of spectral codes, holding also when correcting the previous CHEERS measurements to the latest SPEXACT version (v3.06 -- following the method described in \citealt{mernier2020a}). The most immediate interpretation for such a difference is the peculiar environmental conditions of this galaxy. Likely, ram-pressure stripping -- seen in earlier works \citep[e.g.][]{machacek2005,su2017a,su2017b} and through the cooler SE gas tail in Fig.~\ref{fig:mapsI} has significantly contributed to erode metals from its initial central build-up. If true, this picture provides strong evidence that ram-pressure stripping is an effective mechanism to eject metals from individual galaxies into their cluster environment. Interestingly, NGC\,1404 is also known for showing credible signs of higher turbulence than in most (if not all) other studied systems \citep{pinto2015,ogorzalek2017}, despite its lack of recent AGN activity. A high turbulence in NGC\,1404 would be expected to stir metals, hence to broaden its central metal core. That intuitive picture contrasts instead with the remarkably narrow Fe peak reported here. This suggests that mixing of metals through their hot atmospheres is considerably more sensitive to ram-pressure stripping rather than to turbulence and small-scale gas motions. This does not preclude at all, however, that in more common conditions the kinetic feedback mode of central AGNs dominates the metal distribution process, for instance via jets and buoyant bubbles displacing directly central low-entropy metal-rich gas to higher altitudes \citep{simionescu2008,kirkpatrick2011,kirkpatrick2015}. This would, in fact, naturally connect the narrow extent of the central peak with the absence of (visible) AGN feedback in this galaxy \citep[see also][]{rebusco2006}.

We also note that, like in earlier work \citep[e.g.][]{mernier2017}, the central Fe and Mg peaks are significantly lower than those predicted by the state-of-the-art simulations in ETGs; e.g. the MACER simulations \citep{pellegrini2020} and the MrAGN runs of \citet{choi2020}. The precise origin of these disagreements still has to be determined, but should be (at least partly) related to the complex subgrid physics at play here.

\subsubsection{Two central metal peaks?}

In addition to the central metal peak, the residual maps from Figs.~\ref{fig:mapsII} and \ref{fig:mapsIII} show a secondary peak that is marginally detected with $>$1.2$\sigma$. Intriguingly, this peak is not associated with any particular feature in X-ray/radio/optical SB, nor with particular gas temperature structure. If real, interpreting this peak is not easy, however the least unlikely scenario would be that of a localised region of star formation.

Assuming a line-of-sight velocity of 522~km\,s$^{-1}$ for NGC\,1404 towards NGC\,1399 \citep{su2017a} and, very roughly, that this offset peak is widely entrained by such velocity via ram-pressure stripping, we estimate that metals must have been released less than $\sim$6~Myr in this region before being dislocated. Comparing this time-scale with the ($\sim$one order of magnitude) longer time-scale for massive stars to eject metals via SNcc explosions, it comes that such a hypothetical star-forming region should have remained active and largely visible in blue optical bands. Since, again, no counterpart is visible in the optical nor radio bands, the such star-forming region seems implausible. One should keep in mind, however, the rough assumptions considered here: non-negligible gas viscosity and/or relative steadiness of the inner parts of the galaxy may very well keep this secondary peak essentially intact.

\subsubsection{Metal-poor arc inside the cold front?}

While our (3T) azimuthally averaged Fe profile favours a smooth decrease from $\sim$1 to $\sim$0.6 Solar, our corresponding 2D analysis shows a significant metal-poor arc that is found to lie \textit{inside} the NW cold front. This finding contrasts with other merging/sloshing systems, in which metallicity drops to lower values \textit{outside} their associated cold front \citep[e.g.][]{simionescu2010,ghizzardi2014,werner2016,urdampilleta2019}. To our knowledge, this is the first time that such opposite situation is reported.

This peculiar feature is difficult to interpret. One possibility is that the gas outside the central metal peak was initially enriched at $\sim$0.3~Solar and remained so until the galaxy began to interact with the Fornax ICM. If the ICM and NGC\,1404's atmosphere do not mix well across the merging cold front (e.g. shielded by magnetic draping, whereas in the opposite direction the tail helps mixing the gas efficiently), a metal-poor arc inside the cold front may have survived from the galaxy's motion through the ICM. In a similar scenario, the ICM surrounding NGC\,1404 may have compressed (but not mixed with) the whole metal gradient of the system when interacting with the Fornax cluster. These two possibilities, however, are not supported by numerical simulations, as the latter typically predict the emergence of gas motions inside the cold front as well \citep{heinz2003,markevitch2007}.

Another -- perhaps more likely -- possibility is that the metal-poor arc is not real. Extending the ``double Fe bias'' case discussed in Sect.~\ref{sec:profiles}, projection effects along the cold front region may very well result in a temperature distribution more complex than our simple 3T assumption. In turn, our improved modelling approach may still not be sufficient to provide fully unbiased Fe abundances. Although deriving information from even more complicated multitemperature modelling is challenging at this temperature regime, we note that this cold-front region constitutes a target of choice for future observation campaigns at higher spectral resolution; particularly with \textit{XRISM}/Resolve and, at improved spatial resolution, \textit{Athena}/X-IFU.

\subsubsection{The (dissimilar) Mg and Si spatial distributions}

Our results show that Mg follows Fe remarkably well locally in the hot gas of NGC\,1404. Given that Fe and Mg originate from different SNe types (respectively, SNIa and SNcc) with, in principle, different enrichment histories and time delays, finding a spatially uniform Mg/Fe distribution suggests that these two channels had enriched the hot atmosphere of NGC\,1404 early on, well before the dynamics of the system affected the distribution of these elements (via e.g. ram-pressure and internal motions). This picture, discussed more extensively in the next sections, can be naturally put into the context of the early enrichment scenario for clusters and groups in general, as such spatial uniformity of $\alpha$/Fe ratios has been already reported in the ICM of more massive systems and interpreted in a similar manner \citep{deplaa2006,simionescu2009,simionescu2015,ezer2017,mernier2017}.

More surprising is the Si-rich arc, corresponding to a radial increase of the Si/Fe ratio to super-Solar levels reported between $\sim$2 and 7~kpc (Fig.~\ref{fig:profiles} bottom right-hand panel). To investigate this further, we have refitted spectra from EPIC MOS, EPIC pn, and ACIS sectors extending in the E and W directions, as shown in Fig.~\ref{fig:NGC1404_zoomed}. Interestingly, we find a super-Solar enhancement at $>$1$\sigma$ in \textit{both} directions with similar trends for the three instruments independently, meaning that the Si-rich arc might as well be azimuthally independent. This Si-rich \textit{ring}, confirmed independently by three instruments and two fitting methodologies (i.e. narrow-band in the EPIC profiles, full-band in the EPIC maps and the EPIC/ACIS sectors discussed here), cannot be attributed to instrumental effects only. In fact, the instrumental Si K$\alpha$ line at $\sim$1.75 keV is present in MOS (and likely in ACIS as well) but \textit{not} in pn. 

Such a finding is difficult to explain within any standard SNIa/SNcc enrichment scenario. Indeed, since both Mg and Si are presumably synthesized by SNcc \citep[with a non-negligible SNIa contribution for Si; e.g.][]{mernier2016b,simionescu2019a}, one would naturally expect to see similar spatial distributions for the two elements. If a ring of enhanced SNcc enrichment was indeed in place in NGC\,1404, such a feature should appear at least as prominently in the Mg/Fe ratio. Alternative astrophysical explanations remain sparse. One could speculate either towards a separate enrichment channel for Si (possibly occurring at a distinct epoch, leading to segregated motion/diffusion histories in the hot gas), or towards selective interactions of Si between two phases -- for instance an efficient central depletion of Si into dust, as already proposed to explain metal drops in other systems \citep[e.g.][]{panagoulia2015,mernier2017,lakhchaura2019,liu2019}. We note, however, the remarkably low dust mass estimates that are reported for NGC\,1404 so far \citep{skibba2011,remy2014}. 

Interestingly, this is not the first time that a non-flat Mg/Si trend is reported: using deep \textit{Chandra} observations, \citet{million2011} found a centrally peaked Si/Fe distribution in the core of M\,87, together with a flatter Mg/Fe profile. While the authors interpret their results as an incomplete spectral modelling, in our case we rather speculate on a fitting bias related to the complex temperature structure below the cold front region. If, as discussed before, the absolute Fe abundance is indeed biased low in that region (and so is the Mg abundance, as its line is not entirely dissociated to the Fe-L complex), the Si/Fe may be overestimated. Unlike the Fe-L complex, the Si-K line lies in fact in a band that is likely less sensitive to subtle multitemperature effects. Whether this Si-rich arc/ring is genuine or not will be likely verified by future high-resolution observations (\textit{Athena}/X-IFU and, to some extent, \textit{XRISM}/Resolve).

\subsection{The chemical composition of NGC\,1404}\label{sec:discu_composition}

\begin{figure}
        \centering
                \includegraphics[width=0.49\textwidth, trim={0.5cm 0cm 0cm 0.1cm},clip]{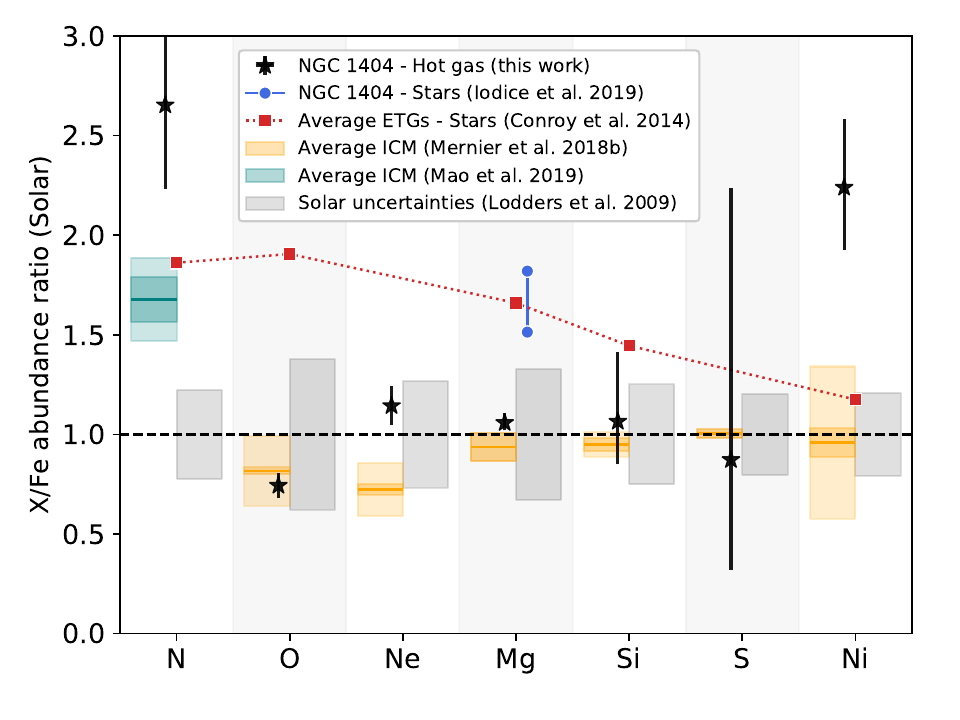}
        \caption{Comparison between the abundance ratios presented in this work (core region) and other measurements, namely: (i) the stellar Mg/Fe ratio derived by \citet{iodice2019} for the same galaxy (delimited by the measurements within 0.5 $R_e$ and beyond 0.65 arcmin); (ii) average stellar X/Fe ratios for a sample of ETGs with similar stellar mass \citep[][$\log M_* = 11.07$]{conroy2014}; and (iii) average X/Fe measured in hot atmospheres of the CHEERS sample \citep[44 clusters, groups, and ETGs;][]{mernier2018b} and in a subsample of eight systems \citep[mostly ETGs;][]{mao2019}. For comparison we also show the Solar uncertainties \citep{lodders2009}.}
\label{fig:ratios_discussion}
\end{figure}

In Fig.~\ref{fig:ratios_discussion}, we compare the abundance ratios obtained conservatively in this work (black stars) with earlier relevant results from the literature. We find an excellent agreement with the abundance pattern of the average ICM \citep[yellow boxes -- the 44 nearby systems of the CHEERS sample, though strongly weighted towards clusters;][]{mernier2018b}. In the same figure, our (gas-phase) ratios are also compared with corresponding stellar ratios from previous work: (i) the stellar ratios as measured with SDSS in a sample of ETGs \citep{conroy2014}\footnote{NGC\,1404 is in fact measured with $M_* = 1.3 \times 10^{11}$~$M_\odot$ \citep{iodice2019}, i.e. close to the highest stellar mass bin of the SDSS sample (red data points on the figure)}; (ii) the stellar Mg/Fe ratio from MUSE observations of NGC\,1404 within the Fornax3D project \citep[][blue limits tracing two aperture radii]{iodice2019}. Overall, it comes that the galaxy's stellar population and its hot atmosphere have a different chemical composition, with a clear $\alpha$/Fe enhancement in the former.

Three ratios deserve further attention, as they do not formally agree with the ICM abundance pattern.

The Ni/Fe ratio is measured to be super-Solar, at variance with the up-to-date ICM estimates \citep{hitomi2017,mernier2018b,simionescu2019a}. However, unlike the hotter ICM -- in which Ni is measured via its K-shell transitions around 7.8~keV, the cooler temperature regime of ETGs makes Ni measurements possible only via the Ni~XIX transitions that are drowned into the Fe-L complex. Even at the RGS energy resolution, these lines are broadened instrumentally and cannot be fully resolved. For instance, the most prominent Ni line in our RGS spectra (at $\sim$14.1\AA) is blended with the Fe~XVIII line (Fig.~\ref{fig:RGScomb}), the latter accounting for $\sim$95\% of the total flux at this wavelength range. Thus, the Ni/Fe best-fitting abundance shown here predominantly reflects subtle fitting adjustments in our RGS spectra (due to e.g. imperfect calibration or instrumental broadening), and therefore should be interpreted with extreme caution.

The N/Fe ratio is measured $>$3$\sigma$ higher than its Solar value. Unlike Ni, the main N line at ETG-like temperatures is well resolved (N~VII at $\sim$25~\AA), hence its abundance measurement should be robust. In fact, super-Solar N/Fe ratios are a rather common feature to hot atmospheres, as already observed in other sources (e.g. the Centaurus cluster -- \citealt{sanders2008}; NGC\,4636 -- \citealt{werner2009}) and in stacked samples as well \citep{sanders2011,mao2019}. The traditional interpretation is that N is released by AGB stars, thus independently of the SNIa and SNcc channels responsible for the enrichment of all the other probed elements. Interestingly, these ratios are at first order consistent with the typical stellar N/Fe ratios of ETGs of similar masses. This may suggest that, unlike the early SNIa/SNcc enrichment discussed above and further, AGB stars constitute a still ongoing (and thus fully decoupled) channel of metal production and release. 

The Ne/Fe ratio agrees well with its Solar value, but is somewhat in tension with (in fact, higher than) previous ICM measurements. An interesting explanation may be the difference of temperature structure between the cool-core ICM and the hot atmosphere of NGC\,1404. Whereas the former undergoes strong temperature gradients -- likely resulting in a complex (projected) temperature distribution, the latter remains essentially isothermal in its very core (Sect.~\ref{sec:profiles}). This difference is important because hotter temperatures naturally boost the emissivity of Fe~XXI and Fe~XXII lines within 12--13~\AA, making the Ne~X line undistinguishable from those. Consequently, the Ne abundance is degenerate with the relative amount of hot vs. cool gas, and both parameters cannot be well disentangled in RGS spectra of cool-core systems. As this is \textit{not} the case here, our Ne/Fe ratio can be considered as better constrained and more reliable\footnote{As a quantitative example: for an NGC\,1404-like gas (i.e. $kT = 0.7$~keV), SPEXACT v3.06 predicts the Ne X line emissivity alone to be more than 3 times higher than the sum of the abovementioned Fe~XXI and Fe~XXII lines. For a $kT = 1$~keV gas, this ratio dramatically drops to 0.4.}. Even better: assuming the Ne/Fe does not change (much) from system to system, our present ratio can be used as a reliable reference to be assumed in all future X-ray cluster/group studies. In turn, fixing Ne/Fe to $\sim$1.13 (or Ne/O to $\sim$1.54 in RGS spectra) could be used to alleviate the above degeneracy and better constrain the temperature structure in other systems.

\subsection{Can the stellar population of NGC\,1404 produce the metals of its hot gas?}\label{sec:discu_budget}

Whereas the mass of metals observed in the ICM and that available from galaxy stellar populations have been already investigated in clusters \citep[e.g.][]{degrandi2004,simionescu2009,ghizzardi2021}, such budgets have not been quantified yet, to our knowledge, in the case of ETGs, particularly with low gas accretion and little AGN feedback. Although not isolated, NGC\,1404 is an excellent candidate for such a case, as its rapid motion towards NGC\,1399 prevents it from accreting external metals. 

The (hot gas-phase) metal mass of a system enclosed within a radius $R$ for an element X can be calculated as:
\begin{equation}\label{eq:enclosed_metal_mass}
M_\mathrm{X}(<R) = \int_0^R \mathrm{X}_\odot \mathcal{X}_\mathrm{H} A_\mathrm{X} \mathrm{X}(r) M_\mathrm{gas}(r) dr,
\end{equation}
or, in its discrete form, as:
\begin{equation}\label{eq:enclosed_metal_mass_discr}
M_\mathrm{X}(<R) = \sum_{r=0}^R \mathrm{X}_\odot \mathcal{X}_\mathrm{H} A_\mathrm{X} X(r) M_\mathrm{gas}(r),
\end{equation}
where $\mathrm{X}_\odot$ is the reference Solar abundance of element X (i.e. the number of X atoms over the number of H atoms), $\mathcal{X}_\mathrm{H}$ is the proto-solar mass fraction of H (adopted as 0.74), $A_\mathrm{X}$ is the atomic weight of element X, and X$(r)$ is the (hot gas-phase) abundance of element X at radius $r$. Moreover, the gas mass profile $M_\mathrm{gas}(r)$ is calculated as:
\begin{equation}\label{eq:gas_mass}
M_\mathrm{gas}(r) =  \mu m_\mathrm{H} n_\mathrm{tot}(r) V(r),
\end{equation}
where $\mu$ is the mean molecular weight of ionized gas (adopted as 0.62), $m_\mathrm{H}$ is the hydrogen mass, $n_\mathrm{tot}(r)$ is the total (i.e. electron + proton) gas density profile, and $V(r)$ is the volume of the considered emitting shell. We use the publicly available\footnote{https://github.com/jeremysanders/dsdeproj} \texttt{dsdeproj} tool described in \citet{russell2008} to deproject each EPIC annulus and to infer the gas density, which can be injected in Eq. (\ref{eq:gas_mass}) assuming spherical shell volumes. For simplicity, all deprojected spectra are fitted with a 2T model, except the 5th and the 9th shells in which a 1T model was necessary to obtain stable fits. We also fix the abundance values to their projected counterparts. In turn, the enclosed gas mass
\begin{equation}\label{eq:enclosed_gas_mass}
M_\mathrm{gas}(<R) = \int_0^R M_\mathrm{gas}(r) dr
\end{equation}
and the enclosed metal mass $M_\mathrm{X}(<R)$ defined above can be derived. The uncertainties on these quantities are estimated via a standard Monte Carlo approach for normal distributions of X$(r)$ and $n_\mathrm{tot}(r)$ parameters following their 1$\sigma$ uncertainties in each annulus/shell $r$. 

\begin{figure}
        \centering
                \includegraphics[width=0.49\textwidth, trim={0cm 0cm 0cm 0cm},clip]{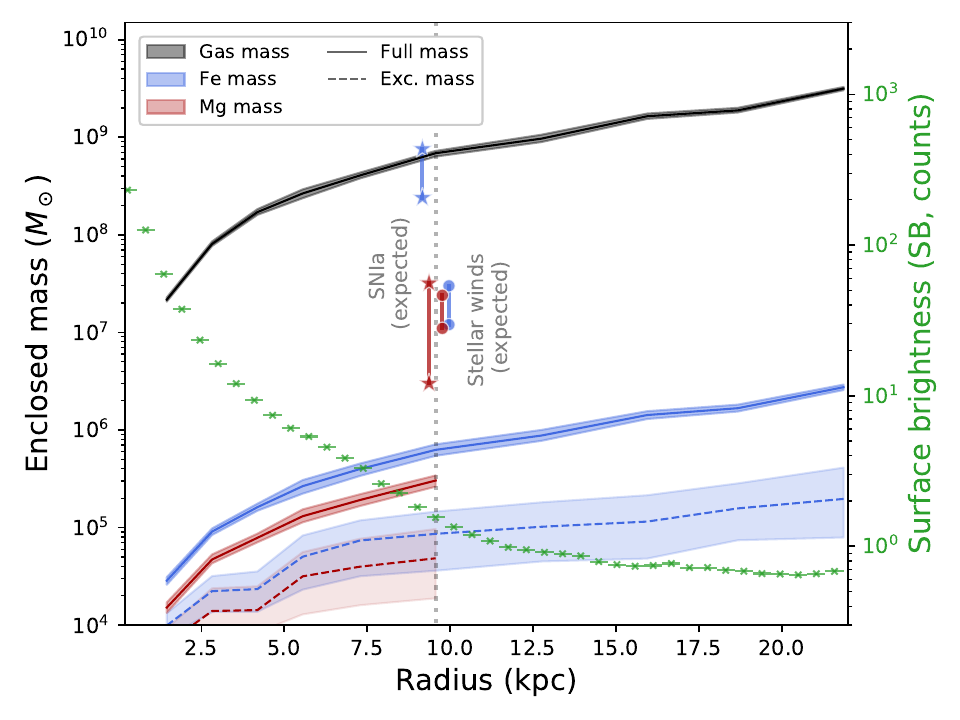}
        \caption{Encircled X-ray gas and metal (Fe and Mg) masses integrated within a given (azimuthally averaged) distance from the centre of NGC\,1404. The solid and dashed curves indicate respectively the fully integrated and excess masses. The vertical line segments, coded with corresponding colours, indicate our estimated ranges of metals produced by SNIa and by stellar winds (see the text). The dotted vertical line corresponds to the radius within which the gas is assumed to have not interacted yet with the Fornax cluster (see Fig.~\ref{fig:profiles}, bottom left-hand panel).}
\label{fig:metal_mass}
\end{figure}

Figure~\ref{fig:metal_mass} shows our estimates of $M_\mathrm{gas}(<R)$, as well as $M_\mathrm{Fe}(<R)$ and $M_\mathrm{Mg}(<R)$ as ``pure'' products of SNIa and SNcc respectively (Si being produced by both channels). Evidently, accounting for the surrounding ICM beyond the cold front radius leads to a monotonic increase of all masses. Restricting our estimates to the gas extent of the galaxy (9.6 kpc), we find a total gas mass of $M_\mathrm{gas} = (6.9 \pm 0.5) \times 10^8$~$M_\odot$ as well as Fe and Mg masses of $M_\mathrm{Fe} = (6.2 \pm 0.8) \times 10^5$~$M_\odot$ and $M_\mathrm{Mg} = (3.0 \pm 0.4) \times 10^5$~$M_\odot$.

In addition to the classical enclosed metal masses defined above, we are also interested in estimating the enclosed metal mass \textit{excess}, $M_\mathrm{X, exc}(<R)$, i.e. the metal mass of the central peak alone (subtracting the ``flat'' abundance from the ``total'' abundance profile). This is done easily by replacing X$(r)$ with X$_\mathrm{exc}(r) = \mathrm{X}(r) - X_0$ in Eqs. (\ref{eq:enclosed_metal_mass}) and (\ref{eq:enclosed_metal_mass_discr}), with $X_0$ adopted here as the lowest abundance value found in the radial profile of element X. Also shown in Fig.~\ref{fig:metal_mass}, these Fe and Mg excess masses are, respectively, $M_\mathrm{Fe, exc} = (9 \pm 5) \times 10^4$~$M_\odot$ and $M_\mathrm{Mg, exc} = (4.8_{-3.0}^{+4.8}) \times 10^4$~$M_\odot$.

First, let us compare these metal masses with those reasonably produced by SNIa through the galaxy's lifetime. Assuming the SNIa rate of \citet{maoz2017} for cluster galaxies ($N_\mathrm{Ia}/M_* = 5.4 \times 10^{-3}~M_\odot^{-1}$), the stellar mass of NGC\,1404 ($12.7 \times 10^{10}~M_\odot$) provides $6.9 \times 10^{8}$ SNIa explosions. Using the SNIa yields from \citet{seitenzahl2013}, and depending on the assumed yield model, this translates into ejected masses of (2.4--$7.6)\times 10^8~M_\odot$ and (0.3--$3.2)\times 10^7~M_\odot$ for Fe and Mg, respectively. Although it is interesting to note that these SNIa-driven masses can largely account for both the Fe \textit{and} the Mg observed masses, the large difference between the Fe and Mg ejecta produced by SNIa (by more than one order of magnitude) does not reflect the factor $\sim$2 difference between the observed masses of these two elements. Unless the escape (and/or depletion) fraction of Fe and Mg is for some reason highly segregated, a full SNIa origin for \textit{all} the metals in NGC\,1404 is thus highly unlikely.

Unlike SNIa, deriving metal masses produced by SNcc through the galaxy's history is difficult as it strongly depends on its star formation history. Quantifying the latter is challenging, although like other massive ETGs NGC\,1404 had likely undergone a short and brief star formation peak around $z \sim 3$ \citep[][see also our discussion in the next section]{thomas2010,yates2013}. It is interesting to note, nonetheless, the absence of current star formation measured by MUSE in NGC\,1404 \citep{iodice2019}. The observed 1.4~GHz radio power of $(1.03 \pm 0.15) \times 10^{19}$~W\,Hz$^{-1}$ \citep{grossova2022} in the core of the galaxy could be explained by star formation that is three times lower than the FUV inferred estimate of $(3.58 \pm 0.04) \times 10^{-2}$~$M_\odot$\,yr$^{-1}$. Assuming this rate to be constant since $z \sim 3$ along with a Salpeter IMF, negligible metal mass from SNcc contribution is expected during this recent epoch.

Finally, let us estimate the metal mass ejected from stars via stellar winds. Assuming a Salpeter initial mass function \citep{salpeter1955}, the wind-driven stellar mass-loss in ETGs at a given time $t$ can be estimated as \citep{ciotti1991}:
\begin{equation}\label{eq:wind_loss}
\dot{M_*}(t) = 1.5 \times 10^{-11} L_B \left(\frac{t}{15 \times 10^9~\mathrm{yr}}\right)^{-1.3},
\end{equation}
with $L_B$ the B band luminosity (in solar $L_{B,\odot}$ units). Adopting $L_B = 2.24 \times 10^{10}~L_{B,\odot}$ \citep{ellis2006} and integrating the above equation between $z=3$ and $z=0$, we find a total mass loss of $12.0 \times 10^9~M_\odot$. Then, adopting the stellar abundance ranges in NGC\,1404 \citep[][]{iodice2019}, we find total stellar wind masses of (1.2--$3.0) \times 10^7~M_\odot$ for Fe and (1.1--$2.4) \times 10^7~M_\odot$ for Mg. This is at least, respectively, $\sim$20 and $\sim$40 times the \textit{total} Fe and Mg masses as observed in the whole galaxy, and two orders of magnitude above the central metal excesses alone. Although the above calculations are rough and should be considered with caution, we note that metals released from stellar winds exceed the observed metal masses even when we assume a constant Solar profile over the whole galaxy extent (e.g. in the extreme case of an excessively broad peak before its erosion by ram-pressure stripping). On paper, stellar winds are thus largely able to account for the metals observed in the hot atmosphere of NGC\,1404. An enrichment through this channel would naturally explain the remarkable Mg-to-Fe spatial homogeneity reported through this work. The picture, however, is different for (cool-core) clusters, as previous work showed that stellar winds are not enough to enrich their entire central metal peak \citep[][]{simionescu2009}. If metals in the hot gas of clusters, groups, and ETGs share a common enrichment channel \citep[as their overall comparisons seem to suggest;][]{mernier2018a,mernier2018b}, the challenge to unify these different views remains difficult. This is further discussed in the next section.

\subsection{The chemical history of NGC\,1404}\label{sec:discu_history}

In the previous sections, we have (i) compared the abundances in the hot atmosphere of NGC\,1404 with those of other astrophysical sources, and (ii) compared the (Fe and Mg) observed metal masses with those expected from their stellar populations. In this last section, we discuss jointly all these findings with the attempt to derive a plausible chemical history for this particular galaxy.

Interpreting the remarkable similarity between the X/Fe ratios derived in the hot halo of NGC\,1404 and in the ICM of rich clusters, the chemical composition of hot atmospheres does \textit{not} depend on the mass of the system over nearly two orders of magnitude in mass. Our aforementioned core analysis essentially probes the central metal peak. Given the particular position and stripping motion of the galaxy within the Fornax ICM, this peak is \textit{directly} associated to NGC\,1404 itself, since it reflects metals that may have been either recently produced by its stellar population, or retained onto the galaxy for a long time; but certainly not accreted recently. 
On the other hand, the abundance pattern of NGC\,1404's atmosphere is clearly distinct from that of its (current) stellar population. The traditional interpretation for enhanced stellar $\alpha$/Fe ratios in massive galaxies such as NGC\,1404, is that they are the direct signature of the galaxy's star formation history \citep[see discussions in e.g.][]{thomas2010,johansson2012,yates2013,conroy2014,simionescu2019a}. Typically, massive ETGs undergo ``downsizing'', i.e. an early, brief, and intense star formation episode that shuts down quickly after its onset. Consequently, the metal content of their stellar generations is predominantly locked with material from early SNcc enrichment -- i.e. mostly $\alpha$-elements. On the contrary, a significant fraction of SNIa explode later on, and their (mostly Fe-peak) material should end up in higher proportions into ETGs' surrounding atmosphere. These initial considerations lead to three possible gas-phase enrichment scenarios.

First is the scenario of an \textit{ongoing enrichment}, coming essentially from both delayed SNIa (Fe-peak elements) and stellar winds from the current stellar population ($\alpha$-elements), and with no recent contribution from SNcc. As seen in Sect.~\ref{sec:discu_budget}, stellar winds and SNIa could have largely produced metals within this whole galaxy's hot atmosphere. Moreover, such an ongoing enrichment would naturally explain the gas metal peak surrounding the bulk of its stellar population. This scenario, however, fails to explain the Solar gas-phase X/Fe ratios, as it would require a nearly perfect compensation between stellar winds and SNIa enrichment products to reproduce exactly this abundance pattern. Unless a remarkable fine-tune is at play over two orders of magnitude in mass, it also seems difficult to unify this picture with that of the ICM enrichment \citep[which exhibits very similar abundances and ratios, and which is thought to have completed beyond $z \sim 2$; e.g.][]{biffi2018b,mernier2018c}.

A second, more likely scenario is that of an \textit{early enrichment}, having started as early as the downsizing epoch of NGC\,1404 (likely $z\sim 3$) and completed no much later -- around $z \sim 2$. In such a scenario, both the stellar and the hot gas abundances are independently locked in time and do not further evolve. The main sources of this early enrichment would then be SNcc, prompt SNIa, and (to some extent) early stellar winds. Since this scenario is clearly favoured in the case of the ICM enrichment, a conciliation between ETGs and clusters scales arises naturally. It would also explain the remarkably Solar chemical composition found at all radii in these two types of systems (though we remind our surprising finding on Si/Fe in Sect.~\ref{sec:discu_distribution}). Although difficult to quantify accurately, restricting our estimates from Sect.~\ref{sec:discu_budget} to the $2 < z < 3$ epoch only provides stellar wind and SNIa metal masses that can still comfortably account for the observed ones (e.g. in excess of a factor of $\gtrsim$5 even for stellar winds). 

A third, no less interesting possibility would be that of metals being present in the hot gas even \textit{before} the star formation episode, hence likely also before the gas started to get heated to X-ray temperatures. This \textit{pre-enrichment scenario} would then enrich the (not yet virialized) galaxy's atmosphere via SNe explosions from early (probably metal-poor) stars, hence having very little to do with the current stellar population of NGC\,1404. Extended to (massive) galaxies in general, such a scenario would have more profound consequences in our general considerations of the cosmic history and cycle of metals. It would have the advantage to unify the ICM and ETGs enrichment in an even more natural way than the early enrichment scenario, as simulations indeed predict a substantial amount of metals to have enriched clusters before $z \sim 3$ \citep[][]{biffi2017}. It would also fit with the most natural explanation to the metal conundrum mentioned in Sect.~\ref{sec:discu_budget}, in the sense that the metal content of hot atmospheres at all scales would have very little to do with their associated stellar population \citep[see e.g.][]{blackwell2021}. However, such a pre-enrichment scenario would hardly explain the narrow metal peak present in the core of NGC\,1404, and would ignore the whole content of metals produced more recently which, in one way or another, must be present in its galactic atmosphere.

As discussed above, the early enrichment -- and, to some extent, the pre-enrichment -- scenarios are favoured to explain the chemical history of NGC\,1404. No matter the scenario, however, the metal budget of this particular galaxy remains difficult to fully capture. Indeed, following our estimates in Sect.~\ref{sec:discu_budget}, a dominant fraction of metals that should have been released by (past \textit{and} ongoing) stellar winds and SNIa is missing in our X-ray observations. It is evident that a substantial amount of these metals had probably left the galaxy and enriched the Fornax ICM -- which is qualitatively in line with the efficiency of ram-pressure stripping to enrich its surroundings \citep[Sect.~\ref{sec:discu_distribution}, especially after a second passage in the cluster;][]{sheardown2018}. The surprise, however, is that no chemical signature of stellar winds nor recent SNIa (via, respectively, enhanced or low $\alpha$/Fe ratios) is found in the hot gas. On the opposite end of massive clusters, we may therefore be facing an ``inverse metal conundrum'', in which the bulk of gas-phase metals are missing. It is interesting to note that these two condundra might (at least partly) compensate for each other: if these metals missing from individual ETGs are actually accreted onto associated clusters, the metal budget of the Universe may be less problematic than previously thought. 

Clearly, probing metal budgets in other ETGs and massive clusters will be essential to further understand the complete cycle of metals at these large scales. In this respect, spatially resolved high-resolution spectroscopy onboard the future missions \textit{XRISM} \citep{xrism2020} and \textit{Athena} \citep{barret2018} will be crucial.


\section{Conclusions}\label{sec:conclusion}


In this paper, we have taken advantage of deep \textit{XMM-Newton} (276~ks) and \textit{Chandra} (657~ks) observations of the elliptical galaxy NGC\,1404, to probe the chemical content of its hot atmosphere with unprecedented details. Plunging toward the centre of the Fornax cluster likely for the second or third time \citep{sheardown2018}, this galaxy with no visible AGN feedback is known to exhibit a merging cold front at the interface of its atmosphere with the surrounding cluster medium, as well as an X-ray gas tail induced by ram-pressure stripping \citep[e.g.][]{machacek2005,su2017a,su2017b}. Following a careful spectral modelling, the Fe, Si, and Mg abundances were measured independently in the core region of the galaxy using the EPIC MOS, EPIC pn, RGS, and ACIS instruments. Similarly, we have also investigated the spatial distribution of these elements (both radially and in 2D maps) using the EPIC instruments. Our results are summarised as follows.

\begin{itemize}

\item Measured through their X/Fe abundance ratios, the chemical composition of the central region is remarkably consistent with that of the hot gas of galaxy clusters, and with that of our Solar System (the only exception being the supersolar N/Fe ratio, likely originating from AGB stars). This pattern, however, differs from the stellar abundance ratios measured either at similar stellar mass bin \citep{conroy2014}, or directly in NGC\,1404 using MUSE \citep{iodice2019}. We also note that the Ne/Fe ratio, measured robustly here, can be useful to better constrain the multiphaseness of the gas, which is otherwise hardly accessible in the (often unresolved) shape of the Fe-L complex.

\item We find a central metal peak significantly narrower than what is typically found in other ETGs and galaxy groups. This may suggest that ram-pressure stripping is an effective process to erode metal peaks in galaxies and eventually dilute elements in the surrounding ICM. 

\item We report a new fitting bias, namely the ``double Fe bias'', where even a two-temperature modelling leads to an underestimate of the Fe abundance in the case of complicated temperature structures. Taking a three-temperature modelling into account, we find for the first time a region of lower metallicity \textit{inside} the NW merging cold front. Since such a metal-poor structure would be difficult to explain within conventional scenarios, we suspect that the abovementioned bias remains at play in this complex region.

\item Unlike the remarkably flat Mg/Fe distribution across the galaxy extent (in line with observations of the hotter ICM/IGrM), we measure a significant increase of the Si/Fe ratio $\sim$3--10~kpc away from the core. This trend is seen independently in EPIC MOS, EPIC pn, and ACIS, hence demonstrating that this is \textit{not} an instrumental bias. This Si-rich ring is difficult to interpret, though it might either reveal complex transfer mechanisms between different phases of the ISM, or relate to the double Fe bias.

\item The mass of metals estimated in the hot gas phase of NGC\,1404 ($M_\mathrm{Fe} \simeq 6 \times 10^5$~$M_\odot$; $M_\mathrm{Mg} \simeq 3 \times 10^5$~$M_\odot$) is found to be 1--2 orders of magnitude lower than the mass of metals expelled by SNIa and stellar winds since $z \sim 3$. Beyond the fact that both sources of enrichment could have largely produced these metals (though the ICM and stellar X/Fe ratios do not favour a sole one), we are witnessing an \textit{inverse} metal conundrum, in which too many metals should have been produced compared to what is currently seen. Nevertheless, the most plausible chemical history for NGC\,1404 is that of an early enrichment (or even pre-enrichment), in which the bulk of metals produced by NGC\,1404 had left the galaxy (and enriched the Fornax cluster) more than 10~Gyr ago, via ancient episode of AGN feedback and/or ram-pressure stripping.

\end{itemize}

This work shows the importance of multiwavelength studies to probe metals in (massive) galaxies at all phases. Future campaigns will be essential in the short-term future to obtain a full coherent picture of the chemical enrichment history, as well as its metal transportation mechanisms from sub-pc scales out to the largest (gravitationally bound) structures of our Universe.

\section*{Acknowledgements}

The authors thank the anonymous referee for insightful comments that helped us to improve this paper. FM thanks Hiroki Akamatsu, Ena Choi, Jelle de Plaa, Simona Ghizzardi, Jelle Kaastra, Anwesh Majumder, L\'{y}dia \v{S}tofanova, Remco van der Burg, Michael Wise, Xioaoyuan Zhang, and Zhenlin Zhu for fruitful discussions and/or complementary data. NW and RG are supported by the GACR grant 21-13491X. AS is supported by the Women In Science Excel (WISE) programme of the Netherlands Organisation for Scientific Research (NWO), and acknowledges the World Premier Research Center Initiative (WPI) and the Kavli IPMU for the continued hospitality. This work is based on observations obtained with \textit{XMM-Newton}, an ESA science mission with instruments and contributions directly funded by ESA member states and the USA (NASA). The scientific results reported in this article are based in part on data obtained from the \textit{Chandra} Data Archive. SRON is supported financially by NWO.

\section*{Data Availability}

The original data presented in this article are publicly available from the \textit{XMM-Newton} Science Archive (https://nxsa.esac.esa.int/nxsa-web/) and in the \textit{Chandra} X-ray Center (https://cda.harvard.edu/chaser/). The respective \textit{XMM-Newton} and \textit{Chandra} data are reduced with the \texttt{SAS} (https://www.cosmos.esa.int/web/xmm-newton/sas) and \texttt{CIAO} softwares (https://cxc.cfa.harvard.edu/ciao) and further analysed spectrally with the \texttt{SPEX} fitting package (https://www.sron.nl/astrophysics-spex). Additional derived data products can be obtained from the main author upon request.



\bibliographystyle{mnras}
\bibliography{NGC1404} 








\bsp	
\label{lastpage}
\end{document}